\documentstyle[preprint,aps]{revtex}
\begin{document}
\title{Point-Form Analysis of Elastic Deuteron Form Factors}
\author{T. W. Allen\footnote{Submitted in partial fulfillment of the requirement
for a Ph.D. in Physics}, W. H. Klink, W. N. Polyzou}
\address{ 
Department of Physics and Astronomy, 
The University of Iowa, 
Iowa City, IA 52242
}
\date{\today}
\maketitle
\footnotetext{P.A.C.S. 03.65.Pm, 21.30.x,21.45.+v, 21.10.Jv,25.10.+s,  25.30.Bf} 
\begin{abstract}
Point-form relativistic quantum mechanics is
applied to elastic electron-deuteron scattering.  The deuteron is
modeled using relativistic interactions that are scattering-equivalent
to the nonrelativistic Argonne $v_{18}$ and Reid '93 interactions.  A
point-form spectator approximation (PFSA) is introduced to define a
conserved covariant current in terms of single-nucleon form factors.
The PFSA is shown to provide an accurate description of data up to
momentum transfers of 0.5 ${\rm GeV}^2$, but falls below the data at
higher momentum transfers.  Results are sensitive to the nucleon form
factor parameterization chosen, particularly to the neutron electric
form factor.
\end{abstract}
\newpage
\narrowtext
\section{Introduction}

Electron scattering is considered to be an ideal tool to study the
electromagnetic structure of hadronic systems.  Relativity
cannot be ignored for momentum transfers that provide information
about the structure of the hadrons at the scale of a few tenths of a
fermi.  In order to understand hadronic systems at this scale,
consistent relativistic models of both the hadronic dynamics and the
hadronic electromagnetic current operator are required.  If the
dynamics and the current operator satisfy cluster properties \cite{sokolov,coest:polyz}, then the
information learned about the structure of the simplest two- and
three-body systems provides the essential components of models needed
to treat complex targets.

Elastic electron-deuteron scattering is the simplest reaction that must
be accurately modeled in order to constrain the dynamical generators
and current operators that are needed to model complex systems.
Because the deuteron is an isoscalar target, it might be expected that
it can be accurately described by a pure impulse
approximation.  Unfortunately, pure impulse approximations are not
consistent with current covariance or current conservation.  An
important goal is to find a physically motivated extension of the
impulse approximation that is consistent with current conservation and
covariance and is also qualitatively consistent with experiment.

In the one-photon-exchange approximation the experimental observables
can be expressed in terms of matrix elements of the hadronic
electromagnetic current operator between the initial and final
eigenstates of the hadronic Hamiltonian.  The general form of these
matrix elements is
\begin{equation}
\langle p'\;j'\;\mu'_j\vert\hat{J}^{\mu}(0)\vert p\;j\;\mu_j\rangle,
\label{matelform}
\end{equation}
where $\vert p'\;j'\;\mu'_j\rangle$ and $\vert p\;j\;\mu_j\rangle$ are
eigenstates of the four-momentum, spin, and three-component of the spin,
in reference frames related by a boost with momentum transfer $Q=p'-p$.
$\hat{J}^{\mu}(0)$ is the hadronic current density at $x=0$.

The dynamical constraints on the current are current conservation,
\begin{equation}
[\hat{P}_{\mu} , \hat{J}^{\mu} (0) ] =0,
\label{curcon}
\end{equation}
and current covariance,
\begin{equation}
U(\Lambda ,a) \hat{J}^{\mu} (x) U^{\dagger}(\Lambda, a) =
(\Lambda^{-1})^{\mu}{}_{\nu} \hat{J}^{\nu}(\Lambda x+a). 
\label{curcov}
\end{equation}
Here $U(\Lambda ,a)$ is the unitary representation of the inhomogeneous
Lorentz group, whose existence is required by relativistic invariance \cite{wigne}. 
$\hat{P}^{\mu}$ is the four-momentum operator, with $U(I,a)=e^{i\hat{P}\cdot a}$.

In applications there are essentially two approaches used to compute
the hadronic current matrix elements. These are the covariant
\cite{gross,blank,logun,wally,humme}
and direct interaction
\cite{carbo,carbo2,lev,lev2,lev3,levap,cckp,frank,schia}
approaches.  Each approach has its own advantages and disadvantages.

Covariant approaches assume that the underlying theory is a local
quantum field theory.  For the case of the deuteron the input is a
covariant current vertex of the form
\begin{equation}
\langle 0 \vert  T(\Psi (x_1)\Psi (x_2) \hat{J}^{\mu} (x) 
\bar{\Psi} (y_1 ) \bar{\Psi} (y_2)) \vert 0 \rangle , 
\label{covcurver}
\end{equation}
which is the vacuum expectation value of a time-ordered product of
nucleon and current fields.
Assuming the existence of an underlying quantum theory, Mandelstam
\cite{mandelstam} showed how to extract the desired current matrix
elements from the vertex.  The Fourier transform of the vertex has
pole terms on the deuteron mass shell.  The residue includes a pair of
Bethe-Salpeter amplitudes and a current matrix element.  The
Bethe-Salpeter normalization condition \cite{itzykson} can be used
with the solution of the homogeneous Bethe-Salpeter equation to remove
the amplitudes from the residue.  What remains is the desired current
matrix element.

Quasipotential equations \cite{gross,blank,logun} are based
on these same concepts, but they introduce constraints designed
to preserve the physical singularities.  Current matrix elements are
extracted using the constrained amplitudes and vertex functions.

Covariant methods are appealing because of their formal connection to a
quantum field theory; however in most applications it is necessary to
model the vertex and Bethe-Salpeter kernel, and to replace the two-point
Green's function with the free two-point function.  Quasi-potential 
methods lead to simpler calculations \cite{wally} than the full 
Bethe-Salpeter approach \cite{humme},  but the reductions complicate 
cluster properties.

Direct interaction approaches attempt to construct consistent models
of $U(\Lambda,a)$ and $\hat{J}^{\mu} (x)$ directly. The transformation
properties require that both $U(\Lambda,a)$ and $\hat{J}^{\mu}(x)$
have an interaction dependence.
Dirac \cite{dirac} addressed the problem of constructing $U(\Lambda,a)$
by including interactions in some of the infinitesimal generators of $U(\Lambda ,a)$.  He
introduced the notion of forms of dynamics which minimize the number of
interaction-dependent generators.  The three main
forms are:   the instant form, where the interactions are in the
Hamiltonian and Lorentz boost generators; the point form,
where the interactions are in the four-momentum; and the
front form, where interactions appear in the operators
that generate transformations transverse to a fixed light front (a
three-dimensional hyperplane tangent to the light cone.)  But while Dirac
identified the different possibilities for putting interactions in selected
generators, he did not show how to actually construct the Poincar\'e
generators with interactions.

The first exact construction of Poincar\'e generators with interactions was due to
Bakamjian and Thomas \cite{bakam} using Dirac's instant form.  There are
Bakamjian-Thomas-like constructions in each of the forms of
dynamics\cite{keist}, and they are scattering equivalent.

While explicit dynamical models of current operators are difficult to
construct, consistent current matrix elements can be obtained by
prescriptions that evaluate selected independent matrix elements using
single nucleon currents.  The remaining current matrix elements can then be
determined by using covariance, current conservation and discrete
symmetries.  These generate the needed dynamical contributions to the 
current matrix elements.

Direct interaction approaches provide an exact treatment of the
symmetries associated with special relativity, but are not
directly related to an underlying field theory.
A number of direct interaction applications to elastic
electron-deuteron scattering exist in the literature.  To date most
applications have used Dirac's instant-\cite{schia}
or front-forms \cite{coest:osteb,cckp,serduke,kondr,teren}
of the dynamics.  The point form of relativistic quantum mechanics has
important simplifying features that are useful in modeling electron
scattering.  The purpose of this paper is to investigate
the hadronic current operator in Dirac's point form of dynamics.

In section II we discuss some of the features of point-form dynamics
and construct a mass operator for the deuteron.  Section III
deals with current operators, their relation to observables, and the point-form spectator approximation.
Then in section IV the numerical results
are discussed and compared with other methods.  Section V presents our conclusions.

\section{Point-Form Relativistic Quantum Mechanics}

Unlike nonrelativistic quantum mechanics, where all the interactions can
be put in the Hamiltonian operator, for relativistic quantum mechanics it
is necessary that at least three generators contain interactions.  This
can already be seen by examining the commutator of the Lorentz boost
generators with the momentum generators.  Such a commutator produces the
Hamiltonian;  if the Hamiltonian contains interactions, then some
combination of boost and momentum generators must also contain
interactions.  
In the instant form additional interactions are put in the boost
generators, leaving the momentum generators free of interactions, while
in the point form the additional interactions are in the momentum
generators, with the boost generators free of interactions.  The front
form puts interactions in a mixture of Lorentz and momentum generators.

Even though all forms of dynamics are scattering equivalent, each has certain
advantages that are useful for specific applications.
The goal of this paper is to analyze elastic deuteron form factors using
the point form.  The point form has a number of features that set it
aside from the other forms.  First, all of the interactions are in the
Hamiltonian and momentum generators, that is, the four-momentum
operator.  Since there are no interactions in the boost or angular
momentum generators, the Lorentz generators are all kinematic and the
theory is manifestly Lorentz covariant.  It is convenient to write
the Poincar\'e commutation relations not in terms of the ten generators,
but rather in terms of the four-momentum operators that contain the
interactions, and global kinematic Lorentz transformations:
\begin{eqnarray}
\label{momcom}
{[\hat{P}_{\mu},\hat{P}_{\nu}]}&=&0;\\
U_{\Lambda}\hat{P}_{\mu} U_{\Lambda}^{-1}&=&(\Lambda^{-1})_{\mu}^{\nu} \hat{P}_{\nu};
\label{momboost}
\end{eqnarray}
where $U_{\Lambda}\equiv U_{\Lambda}(\Lambda,0)$ is a unitary operator representing the Lorentz
transformation $\Lambda$.  These rewritten Poincar\'e relations will be
called the point-form equations, and are the fundamental equations that
have to be satisfied for the system of interest.  The mass operator is
given by $\hat{M}=\sqrt{\hat{P}\cdot\hat{P}}$ and must have a spectrum that is bounded
from below.

Since the interactions are all in the four-momentum operators, which are
the generators of space-time translations, the nonrelativistic
Schr\"odinger equation can be generalized to a Lorentz covariant relativistic
Schr\"odinger equation, namely
\begin{equation}
i\partial\Psi_{x}/\partial x^{\mu}=\hat{P}_{\mu}\Psi_{x},
\label{covrelsch}
\end{equation}
where $x=x^{\mu}$ is the four-vector space-time point.
If the four-momentum operator does not depend explicitly on 
space-time, this equation becomes the eigenvalue equation
\begin{equation}
\hat{P}_{\mu} \Phi=p_{\mu} \Phi.
\label{momeigeq}
\end{equation}
Finally, as will
be shown in the following paragraphs, it is possible to define
states with the property that angular momentum can be coupled in exactly
the same way as is done in nonrelativistic quantum mechanics.

The simplest example of a system satisfying the point form equations is a
one-particle system with mass $m$ and spin $j$.  If $\vert p,\sigma\rangle$ is an
eigenstate of four-momentum $p$ (with $p\cdot p=m^2$) and spin projection $\sigma$, then
\begin{eqnarray}
\label{onepartmom}
\hat{P}_{\mu}\vert p,\sigma\rangle&=&p_{\mu}\vert p,\sigma\rangle\\
U_{\Lambda}\vert p,\sigma\rangle&=&\sum_{\sigma'}\vert\Lambda p,\sigma'\rangle D^j_{\sigma'\sigma}(R_W)
\sqrt{v'_0\over v_0},
\label{onepartboost}
\end{eqnarray}
with $R_W$ a Wigner rotation defined by $R_W=B^{-1}(\Lambda v)\Lambda B(v)$, and 
$B(v)$ a canonical spin (rotationless) boost (see reference \cite{klink2}) with argument $v=p/m$.
$D^j_{\sigma'\sigma}(R_W)$ is a Wigner D function, and the eigenstates are normalized to
\begin{equation}
\langle p',\sigma'\vert p,\sigma\rangle = \delta^3({\bf p}'-{\bf p})\delta_{\sigma'\sigma},
\end{equation}
relativity requiring the $\sqrt{v'_0/v_0}$ factor.

States of many noninteracting particles are tensor products of one-particle states; however
a problem arises when such many-particle states are Lorentz transformed. 
As can be seen from Eq.\ (\ref{onepartboost}) each state is Lorentz transformed by its own
Wigner rotation, which in general are different.  This means that these
multiparticle states cannot be directly coupled together as is the case
nonrelativistically.  Such a problem is resolved by coupling the single-particle states
in the overall rest frame and boosting.  It is convenient to label the state by the system's
four-velocity $v$:
\begin{eqnarray}
\vert v,{\bf k_i},\mu_i\rangle &:=&U_{B(v)}(\vert k_1,\mu_1\rangle...\vert k_n,\mu_n\rangle)\nonumber\\
&=&\sum\left(\vert p_1,\sigma_1\rangle...\vert p_n,\sigma_n\rangle
\prod_i \left[D^{j_i}_{\sigma_i,\mu_i}(R_{W_i})\sqrt{(v'_i)_0\over(v_i)_0}\right]\right),
\label{velstat}
\end{eqnarray}
where $p_i=B(v)k_i$, $\sum {\bf k_i}={\bf 0}$,
and $R_{W_i}=B^{-1}(p_i/m)B(v)B(k_i/m)$.  Under Lorentz transformations,
using the definition, Eq.\ (\ref{velstat}), such velocity states transform as 
\begin{equation}
U_{\Lambda}\vert v,{\bf k_i},\mu_i\rangle =\vert\Lambda v,R_W{\bf k_i},\mu'_i\rangle
\prod_i \left[D^{j_i}_{\mu'_i,\mu_i}(R_W)\sqrt{(v'_i)_0\over(v_i)_0}\right];
\label{veltran}
\end{equation}
where the Wigner rotation $R_W=B^{-1}(\Lambda v)\Lambda B(v)$ is the
same in all the arguments of the D functions and all the internal momentum vectors ${\bf k_i}$. 
That means all the spins as well as the orbital angular momenta can be
coupled together exactly as is done nonrelativistically.  This property
will be used in the following paragraphs for coupling the nucleon spins
together with the relative orbital angular momentum to get the spin of the
deuteron.  From the relation between external and internal momenta, it
follows that the velocity states defined in Eq.\ (\ref{velstat})
are eigenstates of the noninteracting mass operator $\hat{M}_{\rm free}$
and free four-velocity operator $\hat{V}_{\mu}$: 
\begin{eqnarray}
\label{freemass}
\hat{M}_{\rm free}\vert v,{\bf k_i},\mu_i\rangle &=&\sum_i\sqrt{m_i^2+{\bf k_i}^2} \vert v,{\bf k_i},\mu_i\rangle;\\
\hat{V}_{\mu}\vert v,{\bf k_i},\mu_i\rangle &=&v_{\mu}\vert v,{\bf k_i},\mu_i\rangle.
\label{freevel}
\end{eqnarray}

The Bakamjian-Thomas procedure is implemented in the point
form by writing $\hat{P}_{\mu}=\hat{M}\hat{V}_{\mu}$, where now $\hat{M}$ is the sum of free and
interacting mass operators, $\hat{M}=\hat{M}_{\rm free}+\hat{M}_{\rm int}$.  $\hat{M}$ takes the place
of the center of momentum Hamiltonian $\hat{h}=\hat{H}-{\hat{P}^2\over 2M}$ in
nonrelativistic quantum mechanics;  note however that even though there
is only one operator containing interactions, namely the mass
operator, that nevertheless there are interactions in all four components
of the four-momentum operator.

In order that the four-momentum operator satisfy the point-form equations,
Eqs.\ (\ref{momcom},\ref{momboost}), the interacting mass operator must satisfy certain conditions.  To
satisfy Eq.\ (\ref{momcom}), the mass
operator must commute with the four-velocity operator, defined in Eq.\ (\ref{freevel}):
\begin{equation}
\left[ \hat{M},\;\hat{V}^{\mu} \right] =0.
\label{velcommass}
\end{equation} 
This has the consequence that mass and four-velocity can be simultaneously
diagonalized.  Eigenstates of the four-momentum operator can thus be written
as the mass times the four-velocity.  Since the four-velocity is purely
kinematic, it can be factored from the wave function leaving the
covariant Schr\"odinger equation, Eq.\ (\ref{momeigeq}), to become a mass operator eigenvalue
equation,
\begin{equation}
\hat{M}\Phi=\lambda\Phi.
\label{masseigeq}
\end{equation}
Moreover, even though the four-momentum is conserved in reactions, the
total four-momentum is not the sum of the four-momenta of the individual
particles.  Rather what is conserved is the overall four-velocity of the
individual particles, and the mass is then ``off-shell'', not unlike the
situation with Feynman diagrams.  This is to be contrasted with the
instant form, where the three-momentum of all the individual particles
give the total three-momentum of the system, while the energy is ``off-shell''.

The mass operator must also satisfy the other point form equation, Eq.\ (\ref{momboost}),
implying the mass operator is a Lorentz scalar.  On velocity
states this means the kernel of the mass operator must be rotationally
invariant and independent of ${\bf v}^2$, exactly the condition put on nonrelativistic Hamiltonians in
order that they be Galilei invariant.

For a two-body system such as the deuteron, the relevant Hilbert space is
the tensor product of proton and neutron Hilbert spaces, $H=H_p \otimes H_n$.  In
that case the velocity states can be written as
$\vert v,{\bf k},\mu_p,\mu_n\rangle$, where ${\bf k}={\bf k_1}=-{\bf k_2}$, and $\mu_p$ and $\mu_n$ are the
eigenvalues of the three-components of the canonical spins of the proton and neutron respectively.  Because with
velocity states the angular momenta can all be coupled together,
these states can also be written as
$\vert v,|{\bf k}|,j,\mu_j,l,s\rangle$, as in the nonrelativistic case.  The
mass of the two particle state, from Eq.\ (\ref{freemass}), is $2\sqrt{m^2+{\bf k}^2}$; $j$ is
the total angular momentum, while $l$ and $s$ are the orbital and spin
angular momentum respectively.

It is advantageous to express the
interacting mass operator in terms of a mass squared operator with matrix elements:
\begin{eqnarray}
&&\langle v, |{\bf k}|,j,\mu_j,l,s\vert \hat{M}^2_I \vert v',|{\bf k'}|,j',\mu'_j,l',s'\rangle\nonumber\\
&=&
\delta (v - v')\delta_{\mu_j \mu'_j}\delta_{j j'} \langle k,l,s 
\Vert (m^j_I)^2 \Vert k',l',s' \rangle.
\label{masssq}
\end{eqnarray}
A mass operator with a kernel of the form Eq.\ (\ref{masssq}) will satisfy Eq.\ (\ref{velcommass})
and thus the Poincar\'e commutation relations, Eqs.\ (\ref{momcom},\ref{momboost}).
The kernel of $\hat{M}^2_I$ is taken to be
\begin{equation}
\langle k,l,s \Vert (m^j_I)^2 \Vert k',l',s'\rangle := 4m\langle k,l,s \Vert v^j_{nn} \Vert k',l',s'\rangle,
\end{equation}
where $v^j_{nn}$ is a nucleon-nucleon interaction. The mass is then defined by
\begin{equation}
\hat{M}:=\sqrt{\hat{M}^2}; \qquad \hat{M}^2:=4({\bf k}^2+m^2)+\hat{M}^2_I.
\end{equation}
Denoting the eigenvalue of the interacting mass
operator by $\lambda^2$, the equation
\begin{equation}
\hat{M}^2\Phi = (4m^2+4{\bf k}^2+4mv^j_{nn})\Phi = \lambda^2\Phi
\end{equation}
can be rewritten \cite{serduke} in the form of the nonrelativistic Schr\"odinger equation,
\begin{equation}
\left( {{\bf k}^2\over m}+v^j_{nn}\right)\Phi = \left({\lambda^2\over 4m}-m\right)\Phi.
\label{pseudosch}
\end{equation}

This defines a relativistic model of the two-nucleon system.  It can be shown \cite{keist}
that this model leads to a small correction to the nonrelativistic binding energy and has
scattering observables identical to the corresponding nonrelativistic model.  Equation (\ref{masssq})
shows that the solution of Eq.\ (\ref{pseudosch}) leads to simultaneous eigenstates of the
mass, velocity, spin, and z-component of spin.  The Poincar\'e transformation properties of
the deuteron eigenstates are given by
\begin{equation}
\hat{P}^{\mu}\vert v,m_D,j,\mu_j\rangle=\lambda v^{\mu}\vert v,m_D,j,\mu_j\rangle
\end{equation}
and
\begin{equation}
U_{\Lambda}\vert v,m_D,j,\mu_j\rangle=\sum_{\mu'_j}\vert\Lambda v,m_D,j,\mu_j\rangle
D^j_{\mu'_j\mu_j}(R_W(\Lambda,v))\sqrt{(\Lambda v)_0\over v_0},
\end{equation}
where
\begin{equation}
\langle v,|{\bf k}|,j,\mu_j,l,s\vert v',m_D,j',\mu'_j\rangle=\delta(v-v')\delta_{\mu'_j\mu_j}\delta_{j'j}\Psi^j_{ls}(|{\bf k}|).
\label{wavefnref}
\end{equation}
$\Psi^j_{ls}(|{\bf k}|)$ is the nonrelativistic wavefunction associated with one of the two chosen
nonrelativistic potentials.  (There are analogous formulas for the scattering states.)
This provides the desired point-form dynamics.

\section{Current Operators, Form Factors, and Elastic Observables}

The second key element in a theoretical description of electron-scattering 
is a conserved, covariant hadronic current density $\hat{J}^{\mu}(x)$.
In the point form the dynamical Poincar\'e transformations are
the space-time translations. 
Translational covariance can be realized by using the dynamical 
translation operators to
define $\hat{J}^{\mu}(x)$ in terms of $\hat{J}^{\mu}(0)$:
\begin{equation}
\hat{J}^{\mu}(x):=e^{i\hat{P}\cdot x}\hat{J}^{\mu}(0)e^{-i\hat{P}\cdot x}.
\label{jat0}
\end{equation}
The density $\hat{J}^{\mu}(0)$ is assumed to transform as a four-vector \cite{levap,klink}
with respect to the free Lorentz transformations.
 
We now want to define the model current operator in terms of measured one-body current operators.
This is done as follows.  The deuteron matrix elements of $\hat{J}^{\mu}(0)$ are defined in terms of
their Breit-frame values with $Q$ in the ${\hat{\bf z}}$ direction:
\begin{equation}
\langle {\bf Q}/2,1,\mu'_j\vert\hat{J}^{\mu}(0)\vert -{\bf Q}/2,1,\mu_j\rangle.
\end{equation}

For $\mu=0,1,2$ the current matrix elements are defined in terms of the single-nucleon current
matrix elements:
\begin{eqnarray}
&&\langle {\bf Q}/2,1,\mu'_j\vert\hat{J}^{\mu}(0)\vert -{\bf Q}/2,1,\mu_j\rangle=\nonumber\\
&&\langle {\bf Q}/2,1,\mu'_j\vert\left(\hat{J}^{\mu}_p(0)\otimes\hat{I}_n
+\hat{I}_p\otimes\hat{J}^{\mu}_n(0)\right)\vert-{\bf Q}/2,1,\mu_j\rangle.
\label{curmatel}
\end{eqnarray}
Current conservation requires that
\begin{equation}
\sum_{\mu}Q_{\mu}\langle{\bf Q}/2,1,\mu'_j\vert\hat{J}^{\mu}(0)\vert-{\bf Q}/2,1,\mu_j\rangle=0,
\end{equation}
which generates a dynamical contribution $\hat{J}^{\mu}_{pn}(0)$ to the $\hat{\bf z}$ component of the current:
\begin{eqnarray}
&&\langle{\bf Q}/2,1,\mu'_j\vert\hat{J}^3_{pn}(0)\vert-{\bf Q}/2,1,\mu_j\rangle\nonumber\\
&&=-\langle{\bf Q}/2,1,\mu'_j\vert\left(\hat{J}^3_p(0)\otimes\hat{I}_n
+\hat{I}_p\otimes\hat{J}^3_n(0)\right)\vert-{\bf Q}/2,1,\mu_j\rangle.
\end{eqnarray}
These relations define the components of the Breit-frame matrix elements of $\hat{J}^{\mu}(0)$.  The
remaining deuteron matrix elements of $\hat{J}^{\mu}(x)$ are fixed by kinematic Lorentz covariance
and dynamical space-time translational covariance.  Although the current matrix element is defined
in the Breit frame, the expression for the general current matrix element is Lorentz covariant, as can
be seen in reference \cite{klink}, Eq.\ 3.31.

The computation of the matrix elements is carried out by inserting single-particle
intermediate states in the velocity basis that was used to formulate the dynamical model in the
previous section.
The deuteron wavefunction in the basis Eq.\ (\ref{wavefnref}) has the form
\begin{equation}
\langle v,|{\bf k}|,j,\mu_j,l,s|v',m_D,j,\mu'_j\rangle=
\delta(v-v')\delta_{\mu'_j\mu_j}\delta_{j\;1}\delta_{s\;1}[\delta_{l\;0}u_0(k)+\delta_{l\;2}u_2(k)],
\end{equation}
where $u_0(k)$ and $u_2(k)$ are the nonrelativistic S and D state deuteron wavefunctions.
Transformation coefficients \cite{keist,klink2} are used to express this in terms of single-particle
basis elements:
\begin{eqnarray}
&&\langle v_1,\mu_1,v_2,\mu_2\vert v',\mu'_j,m_D\rangle \nonumber \\
&&=\delta^3[{\bf v'}-{\bf v}({\bf v_1},{\bf v_2})]
{\delta[k-k({\bf v_1},{\bf v_2})]\over k^2} \left\vert{\partial({\bf v},{\bf k})\over\partial({\bf v_1},{\bf v_2})}\right\vert^{1/2}\nonumber\\
&&\times D^{1/2}_{\mu_1\mu'_1}[B^{-1}(v_1)B(v)B(k_1)]D^{1/2}_{\mu_2\mu'_2}[B^{-1}(v_2)B(v)B(k_2)]
Y_{l\mu_l}[\hat{\bf k}_1({\bf v_1},{\bf v_2})]\nonumber\\
&&\times\langle{1\over 2},\mu_1,{1\over 2},\mu_2\vert 1,\mu_s\rangle\langle l,\mu_l,1,\mu_s\vert 1,\mu'_j\rangle
\left\{\delta_{l\;0}u_0[k({\bf v_1},{\bf v_2})]+\delta_{l\;2}u_2[k({\bf v_1},{\bf v_2})]\right\}.
\end{eqnarray}
These expressions can be used to compute the current matrix element
\begin{eqnarray}
&&\langle v,m_D,1,\mu_j\vert\hat{J}^{\mu}_{SA}(0)\vert v',m_D,1,\mu'_j\rangle\nonumber\\
&&=\int\langle v,m_D,1,\mu_j\vert v_1,\mu_1,v_2,\mu_2\rangle\langle v'_1,\mu'_1,v'_2,\mu'_2\vert v',m_D,1,\mu'_j\rangle\nonumber\\
&&\times\left[\langle v_1,\mu_1\vert\hat{J}^{\mu}_1(0)\vert v'_1,\mu'_1\rangle\delta^3({\bf v'_2}-{\bf v_2})\delta_{\mu'_2\mu_2}+
\langle v_2,\mu_2\vert\hat{J}^{\mu}_2(0)\vert v'_2,\mu'_2\rangle\delta^3({\bf v'_1}-{\bf v_1})\delta_{\mu'_1\mu_1}\right],
\end{eqnarray}
where the nucleon current matrix elements are given by Eq.\ (\ref{curmatel}).  After integrating out
the delta functions, one is left with a final three-dimensional integral:
\begin{eqnarray}
\langle v',m_D,1,\mu'_j\vert\hat{J}^{\mu}(0)\vert v,m_d,1,\mu_j\rangle
&=&\sum_{\mu'_1\mu_1}\sum_{\mu'_2\mu_2}\sum_{\mu'_s\mu_s}\sum_{l'l}
\sum_{\mu'_l\mu_l}\sum_{\sigma'_1\sigma_1}\int{d^3k}\nonumber \\
C^{1\;1/2\;1/2}_{\mu'_s\mu'_1\mu'_2}C^{1\;1/2\;1/2}_{\mu_s\mu_1\mu_2}&\times&
C^{1\;l\;1}_{\mu'_j\mu'_l\mu'_s}C^{1\;l'\;1}_{\mu_k\mu_l\mu_s} \nonumber\\
\times Y^*_{l'\mu'_l}(\theta',\phi')u_{l'}(|{\bf k'}|)&\times&
Y_{l\mu_l}(\theta,\phi)u_l(|{\bf k}|)\nonumber\\
\times D^{*\;1/2}_{\mu'_1\sigma'_1}\{R^{-1}_W[k'_1,B(v')]\}&
\times&D^{*\;1/2}_{\mu'_2\sigma'_2}\{R^{-1}_W[k'_2,B(v')]\}\nonumber\\
\times D^{1/2}_{\sigma_1\mu_1}\{R_W[k_1,B(v)]\}&
\times&D^{1/2}_{\sigma_2\mu_2}\{R_W[k_2,B(v)]\}\nonumber\\
\times\bar{u}(p'_1\sigma'_1)\{\gamma^{\mu}F_1[(p'_1-p_1)^2]&+&
i\sum_{\nu}\sigma^{\mu\nu}{(p'_1-p_1)_{\nu}\over2m_N}F_2[(p'_1-p_1)^2]\}u(p_1\sigma_1)\nonumber\\
+\{ 1 &\leftrightarrow& 2\};
\end{eqnarray}
where the C's are Clebsch-Gordan coefficients and the conventions for the spinors,
gamma and sigma matrices are those of Bjorken and Drell \cite{bjork}.

In this form it can be seen that the momentum of the unstruck particle (the spectator) is
unchanged, while the struck particle's momentum is changed, but
the impulse given to the struck particle is {\it not} the impulse given to the deuteron.  For this reason we
call this the point-form
spectator approximation (PFSA).  It should not be confused with the use of the term spectator
approximation in, for example, reference \cite{gross}.
The practical advantage of the PFSA is that the steps above can be generalized
to any hadronic target.  Moreover, the current matrix element is generally Lorentz covariant
and can be evaluated in any frame.

Because the interactions in the point form are in the
four-momentum, in the PFSA the momentum transfer seen by the scattered
nucleon is not the same as the momentum transfer seen by the nucleus.
In Appendix A we show that the relationship between the momentum $Q$ transferred to the deuteron
and the momentum transferred to the interacting nucleon is
\begin{eqnarray}
\label{pfq2}
\vert(p'_1-p_1)^2\vert &=& Q^2{4(m^2_N+{\bf k}^2_{\bot})\over m^2_D}(1+{Q^2\over 4m^2_D}) \\
&>&Q^2{4m^2_N\over m^2_D}(1+{Q^2\over 4m^2_D})>Q^2.
\label{biggerq2}
\end{eqnarray}
That is, the point-form momentum transfer seen by an individual nucleon will be greater in magnitude
than the total deuteron momentum transfer $Q^2$.

Two important implications follow from the equations (\ref{pfq2}) and (\ref{biggerq2}) above.  First, the PFSA momentum transfer
depends on the internal momentum ${\bf k}$, which is a variable of integration.  Thus in the PFSA,
form factors depending on $(p'_i-p_i)^2$ must remain inside the integral.  Second, since
$\vert(p'_i-p_i)^2\vert >Q^2$, the deuteron form factors will fall off faster in the point-form
calculations than in forms where $\vert (p'_i-p_i)^2\vert =Q^2$.

The input to the PFSA are single-nucleon current operators.  The general structure of these
operators follows from covariance, parity, hermiticity, and time-reversal symmetry.  For a
spin-1/2 target the conditions imply that all matrix elements can be expressed in terms
of the Dirac form factors, $F_1(Q^2)$ and $F_2(Q^2)$.  The general expression has the form \cite{bjork}
\begin{equation}
\langle p,\nu\vert J^{\mu}(0)\vert p',\nu'\rangle
=\bar{u}_{\nu'}(p')\left[F_1(Q^2)\gamma^{\mu}+F_2(Q^2)
\sum_{\alpha}{iQ_{\alpha}\sigma^{\mu\alpha}
\over 2m}\right]u_{\nu}(p),
\label{obc}
\end{equation}
where $u_{\nu}(p)$ and $\bar{u}_{\nu'}(p')$ are Dirac spinors.
In this form the one-body matrix elements are easily evaluated in any kinematic frame.

The Sachs electric and magnetic form factors of the nucleons are
\begin{eqnarray}
\label{nucge}
G_E(Q^2) &=& \sqrt{1+\tau} \langle {\bf Q}/2,{1\over 2},{1\over 2}\vert\hat{J}^0(0)\vert -{\bf Q}/2,{1\over 2},{1\over 2}\rangle; \\
G_M(Q^2) &=& \sqrt{1+\tau\over\tau} \langle {\bf Q}/2,{1\over 2},{1\over 2} \vert\hat{J}^1(0)\vert
 -{\bf Q}/2,{1\over 2},-{1\over 2}\rangle;
\label{nucgm}
\end{eqnarray}
where $\tau={Q^2/4m^2}$.  (Here the standard frame is the Breit frame, where
the nucleon enters with momentum $-Q/2$ and exits with momentum $Q/2$, both along the z-axis,
which is also the axis along which the spin projection is measured.)  These Dirac and Sachs form factors
are related by
\begin{eqnarray}
\label{nucf1}
F_1(Q^2) &=& {1\over 1+\tau}[G_E(Q^2)+\tau G_M(Q^2)]; \\
F_2(Q^2) &=& {1\over 1+\tau}[G_M(Q^2)-G_E(Q^2)].
\label{nucf2}
\end{eqnarray}
The input we use to define the model PFSA current are the single-nucleon form factor parameterizations
of Gari-Kr\"umpelmann \cite{gari} and Mergell-Meissner-Drechsel \cite{merge}.

The experimental observables for the deuteron and nucleon are well known.
The elastic observables $A(Q^2)$ and $B(Q^2)$ are extracted from the Rosenbluth formula for the
cross-section of unpolarized scattering in the lab frame,
\begin{equation}
{d\sigma\over d\Omega} = {\alpha^2\cos^2(\theta/2)\over 4E^2\sin^4(\theta/2)} {E'\over E}
[A(Q^2)+B(Q^2)\tan^2(\theta/2)],
\label{rosen}
\end{equation}
where $\alpha$ is the fine-structure constant, $\theta$ the scattering angle, and $E$ and $E'$
the initial and final energies.  For the nucleons, it can be shown \cite{gourd} that
\begin{eqnarray}
\label{nuca}
A(Q^2) &=& {G^2_E(Q^2)+\tau G^2_M(Q^2)\over 1+\tau}; \\
B(Q^2) &=& 2\tau G^2_M(Q^2).
\label{nucb}
\end{eqnarray}
For spin-1/2 particles, measurements of $A(Q^2)$ and $B(Q^2)$
suffice to determine $G_E$ and $G_M$.  Various models of the
nucleon form factors \cite{gari,merge,chung,lomon,hoehl} have been constructed.

The deuteron has three independent form factors.  A common classification is to denote them as the charge
monopole $G_E$, magnetic dipole $G_M$, and electric quadrupole $G_Q$ form factors.  As current matrix elements,
these are defined in the Breit frame:
\begin{eqnarray}
\label{deutge}
G_E &=& {1\over 3}\langle {\bf Q}/2,1,0\vert\hat{J}^0(0)\vert -{\bf Q}/2,1,0\rangle \nonumber \\
    &+& {2\over 3}\langle {\bf Q}/2,1,1\vert\hat{J}^0(0)\vert -{\bf Q}/2,1,1\rangle; \\
\label{deutgm}
G_M &=& \sqrt{2\over\eta}\langle {\bf Q}/2,1,1\vert\hat{J}^1(0)\vert -{\bf Q}/2,1,0\rangle; \\
G_Q &=& {1\over 2\eta}[\langle {\bf Q}/2,1,0\vert\hat{J}^0(0)\vert -{\bf Q}/2,1,0\rangle \nonumber \\
    &-& \langle {\bf Q}/2,1,1\vert\hat{J}^0(0)\vert -{\bf Q}/2,1,1\rangle ];
\label{deutgq}
\end{eqnarray}
where $\eta=Q^2/4m^2_D$.
These form factors have the static limits
\begin{eqnarray}
\label{deem}
G_E(0) &=& e; \\
\label{demm}
\lim_{Q^2\rightarrow 0} G_M(Q^2) &=& e{m_D\over m_N}\mu_D; \\
\lim_{Q^2\rightarrow 0} G_Q(Q^2) &=& em^2_DQ_D;
\label{deqm}
\end{eqnarray}
where $e$ is the charge, $\mu_D$ the magnetic dipole moment, and
$Q_D$ the electric quadrupole moment of the deuteron.

The Rosenbluth formula alone cannot determine all three of the deuteron's form factors.
The other independent observable normally measured is the deuteron tensor polarization $T_{20}$, defined as
\begin{equation}
T_{20}:=\sqrt{2}{d\sigma^1-d\sigma^0\over d\sigma},
\label{deutt20}
\end{equation}
where $d\sigma^{\mu}$ refers to the differential cross-section with helicity $\mu$.  Conventionally it
is displayed at a 70${}^{\circ}$ angle in the lab frame.
The deuteron elastic observables are:
\begin{eqnarray}
\label{deo:beg}
A(Q^2) &=& G^2_E+{8\over 9}\eta^2G^2_Q+{2\over 3}\eta G^2_M; \\
\label{deo:mid}
B(Q^2) &=& {4\over 3}\eta(1+\eta)G^2_M; \\
T_{20}(Q^2) &=& -\sqrt{2}\eta {{4\over 9}\eta G^2_Q+{4\over 3}G_QG_E+{1\over 3}fG^2_M\over
  A(Q^2)+B(Q^2)\tan^2(\theta/2)};
\label{deo:end}
\end{eqnarray}
where $f=1/2+(1+\eta)\tan^2(\theta/2)$.

\section{Numerical Results and Comparisons}

One purpose of this work was to test the point-form spectator approximation on the
simplest nucleus, the deuteron, where realistic interactions and nucleon form factors
are available.  Comparisons with other models are then an
indication of the relative size of the required two-body currents.

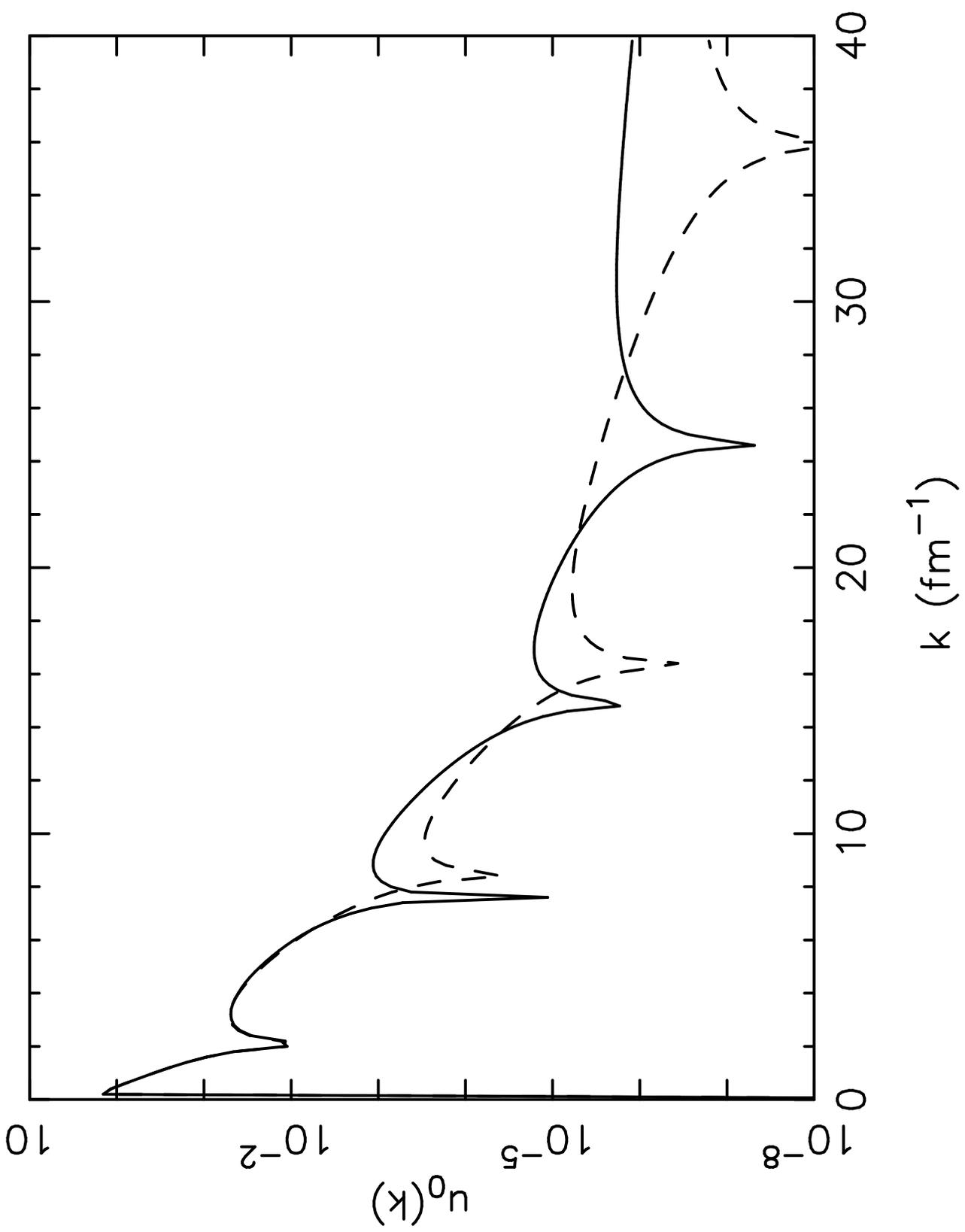
\begin{figure}
\caption{The bound state S (l=0) deuteron wavefunction, using the Argonne (solid) and Reid (dashed) potentials.}
\label{swaveplot}
\end{figure}

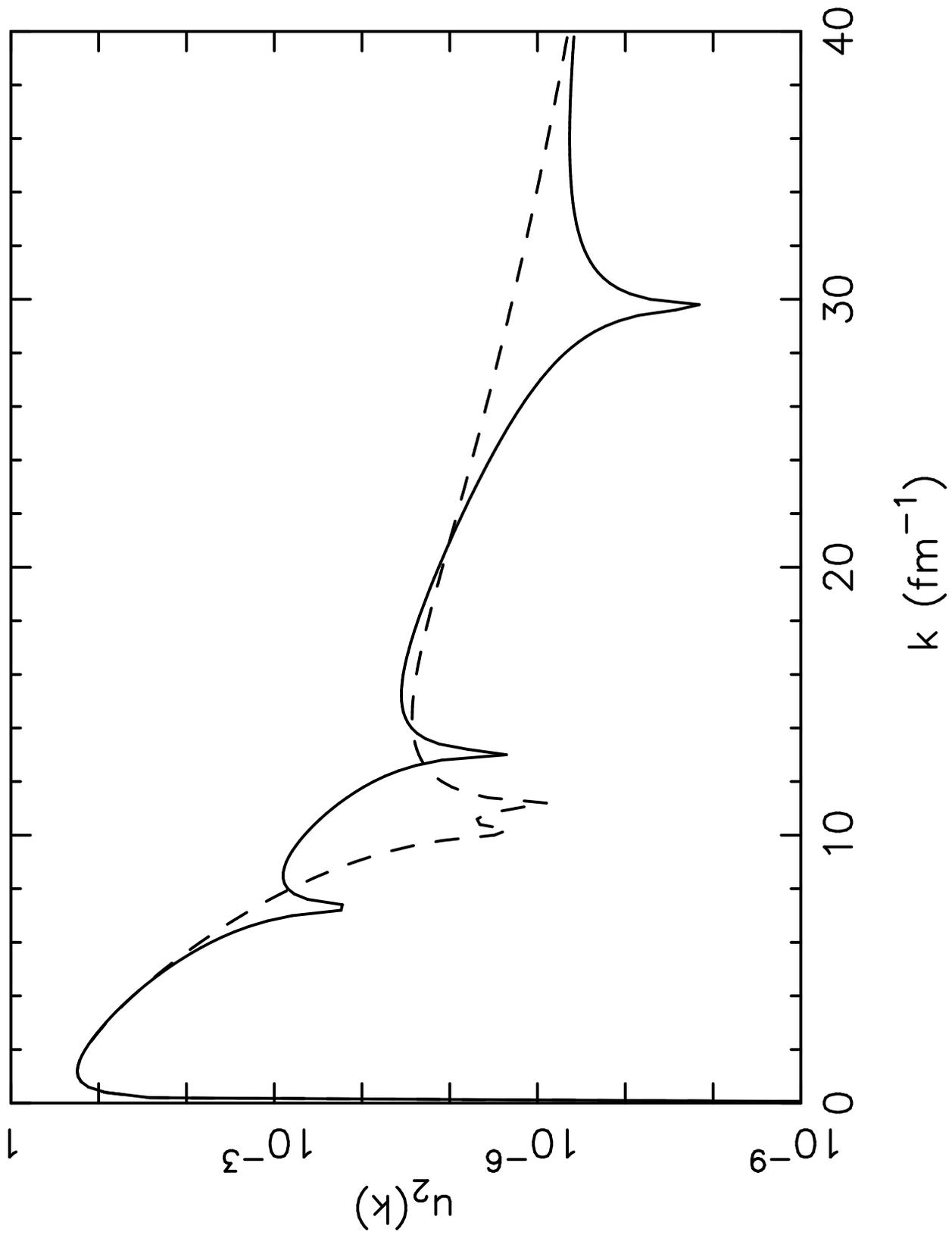
\begin{figure}
\caption{The bound state D (l=2) deuteron wavefunction, using the Argonne (solid) and Reid (dashed) potentials.}
\label{dwaveplot}
\end{figure}

For the nucleon-nucleon interaction, we have used the Argonne $v_{18}$ \cite{argon} and
Reid '93 \cite{reid} potentials to construct a mass operator $\hat{M}$.  The S (l=0) and
D (l=2) state wavefunctions are plotted in Figures \ref{swaveplot} and \ref{dwaveplot}.
The only significant differences between the wavefunctions these potentials produce
in configuration space occur below 0.4 fm for the S wave and below 1.0 fm for the D wave.  In
momentum space the wavefunctions do not exhibit significant differences up to 5 ${\rm fm}^{-1}$,
about 1 GeV, above which they do differ noticeably.  The effects on the choice of interaction may
therefore be expected to be relevant at higher momentum transfers, but as will be seen, these
high-momentum differences in the wavefunction make only slight differences in the calculations.

The PFSA currents are constructed using
the Gari-Kr\"umpelmann \cite{gari} and Mergell-Meissner-Drechsel \cite{merge} parameterizations
of the nucleon form factors.
At the range of momentum transfer under consideration, the parameterizations give very similar results
for the proton form factors and the neutron magnetic form factor.  The neutron electric form factor,
however, varies significantly between the two.

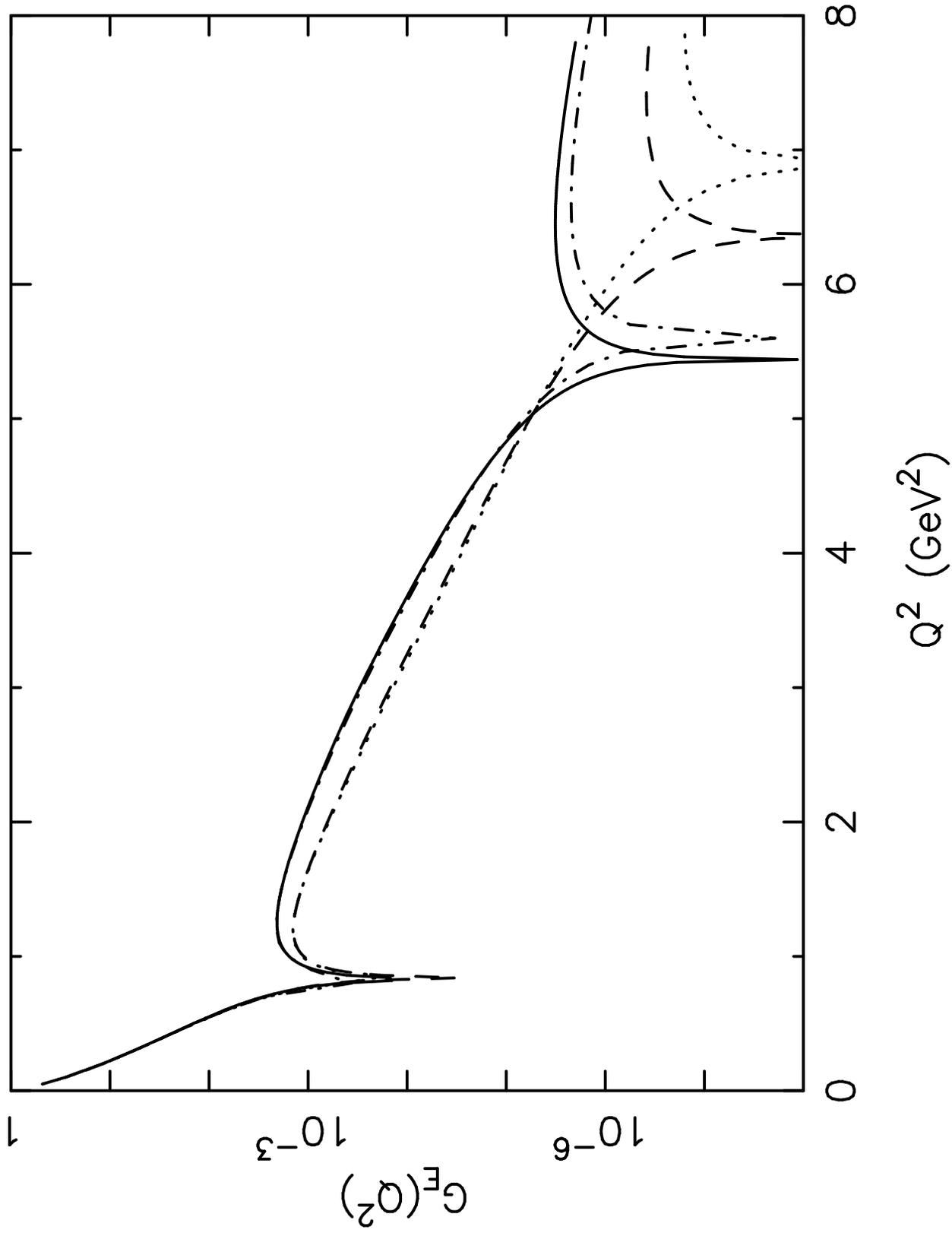
\begin{figure}
\caption{$G_E(Q^2)$ for the Argonne potential with G-K (solid) and MMD (dashed) form factors, and
for the Reid potential with G-K (dash-dot) and MMD (dotted) form factors.}
\label{pagesplot}
\end{figure}

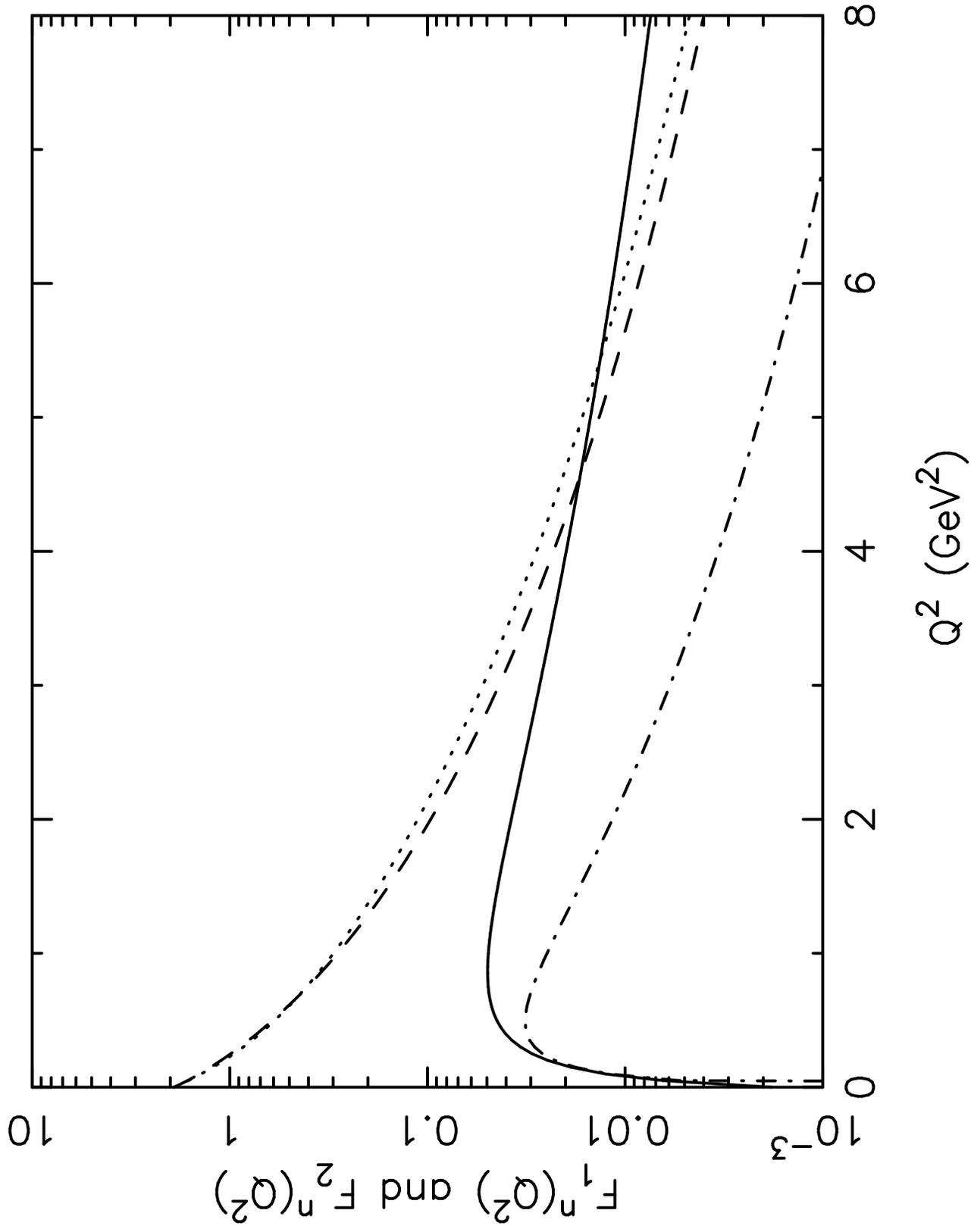
\begin{figure}
\caption{The neutron electric and magnetic form factors of G-K (dotted and dash-dot),
and of MMD (dashed and solid).}
\label{neutplot.ps}
\end{figure}

The deuteron form factor $G_E$ has been calculated using both form factor parameterizations and
both nucleon-nucleon potentials.  The absolute values of the results are displayed in Figure \ref{pagesplot}.
%pagesplot.ps
The results are independent of the nucleon-nucleon potential used, except for small variations
at high momenta.  The primary differences in $G_E$ and $G_Q$ are due to the nucleon form
factor parameterizations.  (Figure \ref{neutplot.ps} compares the G-K and MMD neutron form factors.
Note that the major difference is in the parameterization of the neutron electric form factor;
the neutron magnetic form factor parameterizations, and the proton parameterizations as well, are very similar.)
For $G_E$, both the Gari-Kr\"umpelmann and the Mergell-Meissner-Drechsel form
factors agree at low momentum transfers and have zeroes
near 0.8 ${\rm GeV}^2$.  The G-K form factors predict a second zero near 5.5 ${\rm GeV}^2$
while MMD predicts a second zero between 6 and 7 ${\rm GeV}^2$.  Because the two form factors
are almost identical except for the parameterization of the neutron electric form factor, this
would suggest that the neutron form factor is the dominant cause of the differences in the
calculations of $G_E$.

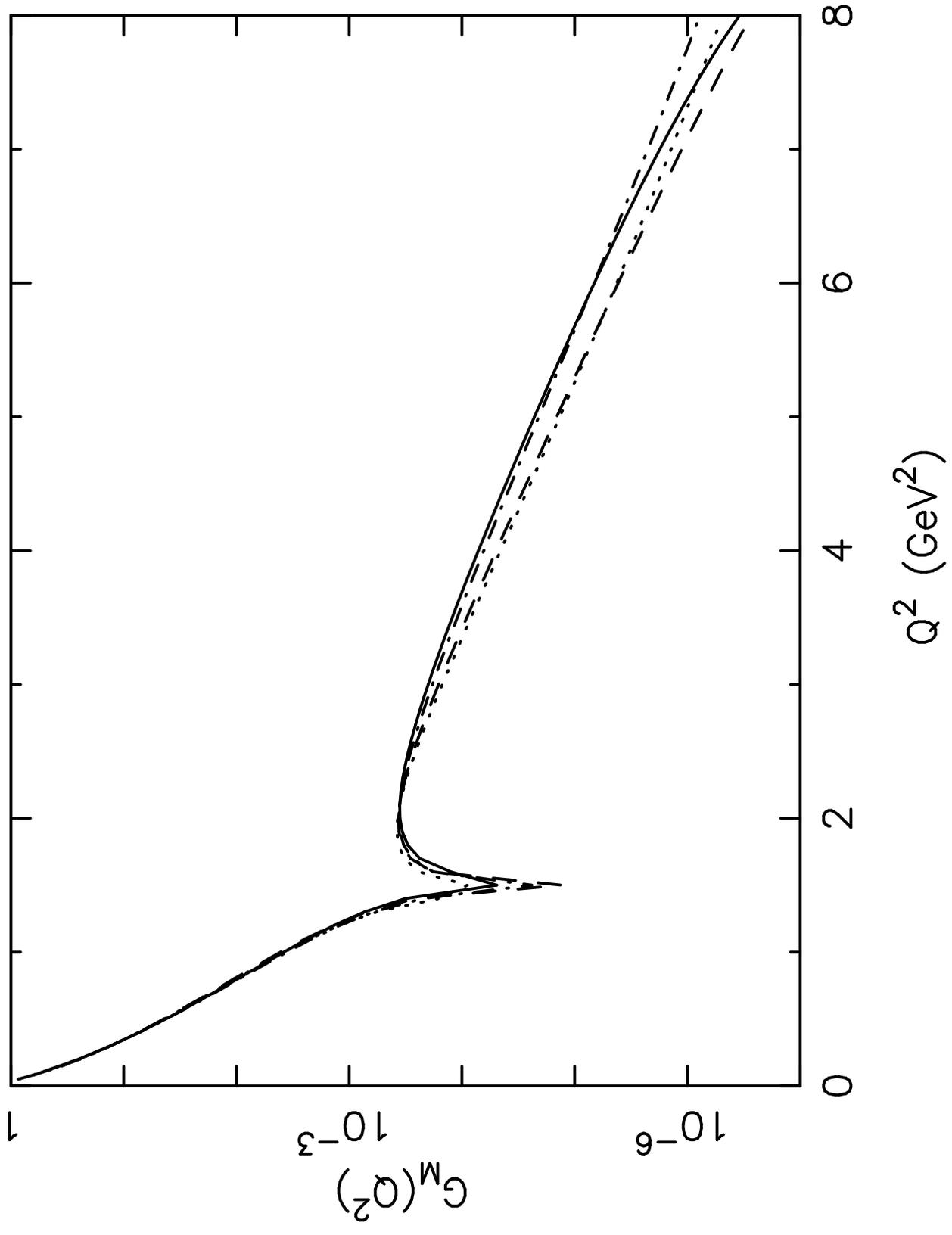
\begin{figure}
\caption{$G_M(Q^2)$ for the Argonne potential with G-K (solid) and MMD (dashed) form factors, and
for the Reid potential with G-K (dash-dot) and MMD (dotted) form factors.}
\label{pagmsplot}
\end{figure}

Figure \ref{pagmsplot} illustrates the dependence of the magnitude of the form factor $G_M$ on the
potential and on the nucleon form factors used.
%pagmsplot.ps
Both parameterizations predict the
same behavior up to the first zero, this time at 1.6 ${\rm GeV}^2$, and within the range studied, fall off
with almost identical behavior.  Comparison with Figure \ref{pagesplot} would suggest that the neutron
electric form factor has little effect on the calculation of $G_M$.  A further comparison to
experimental data can be made by examining the static limit of $G_M$.  Equation (\ref{demm}) relates this
limit to the deuteron magnetic moment, and Table \ref{static} displays the results.
In the static limit, the parameterization of the nucleon form factors does not affect
the results, while the choice of nucleon-nucleon interaction does.  This is
expected, as the form factors must approach precise limits as $Q^2\rightarrow 0$, while the
momentum-space wavefunctions have no such constraints.

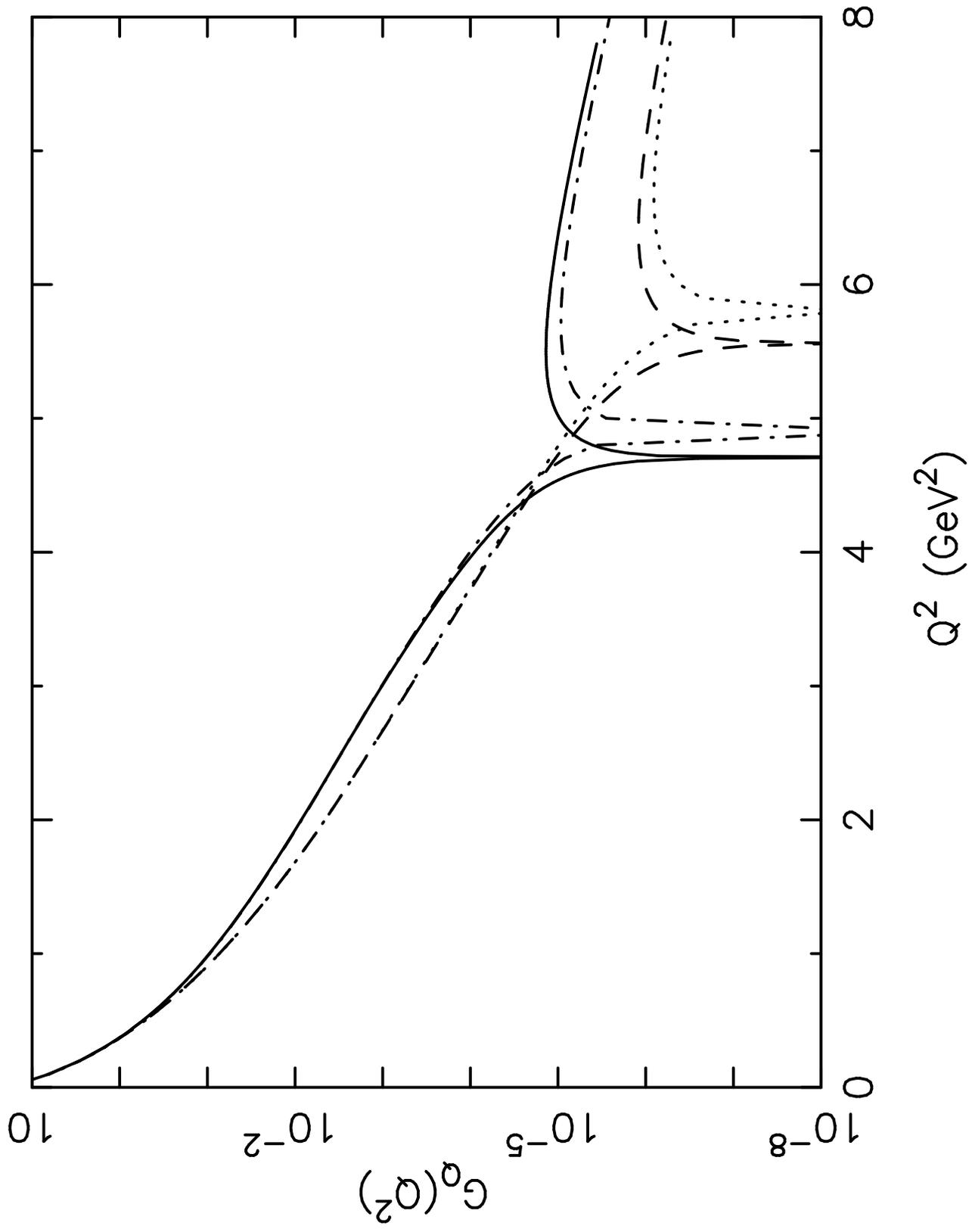
\begin{figure}
\caption{$G_Q(Q^2)$ for the Argonne potential with G-K (solid) and MMD (dashed) form factors, and
for the Reid potential with G-K (dash-dot) and MMD (dotted) form factors.}
\label{pagqsplot}
\end{figure}

This procedure is repeated for $G_Q$ in Figure \ref{pagqsplot}.
%gqsplot.ps
As was the case with $G_E$, there is little difference due to the potential used,
but a noticeable difference between the predictions of the G-K and MMD parameterizations.
The G-K form factors show a zero between 4.5 and 5.0 ${\rm GeV}^2$, while the MMD form
factors produce a zero approximately one ${\rm GeV}^2$ higher.  Further, the magnitude of the
G-K results is greater than that of MMD almost everywhere throughout. The different neutron electric
form factor parameterizations is the primary cause of the differences in the results.
The deuteron electric quadrupole moment (Eq.\ \ref{deqm}), displayed in Table 1, differs from the experimental result,
the calculated values approaching about 90\% of the experimental value, as opposed
to 99\% for the magnetic moment calculations.  This is consistent with other models
\cite{lev,schia,lomon,argon,kobus,carls,ckc}.

To summarize, these point-form calculations imply that the deuteron form factors are essentially
independent of the potential (Argonne $v_{18}$ or Reid '93) used, but depend more significantly
on the parameterizations of the form factors, and in particular on the neutron electric
form factor, as this is the only substantial difference between the G-K and MMD parameterizations.
The static moments are similar to predictions in other realistic models
with the electric quadrupole moment differing with experiment by about 10\%.

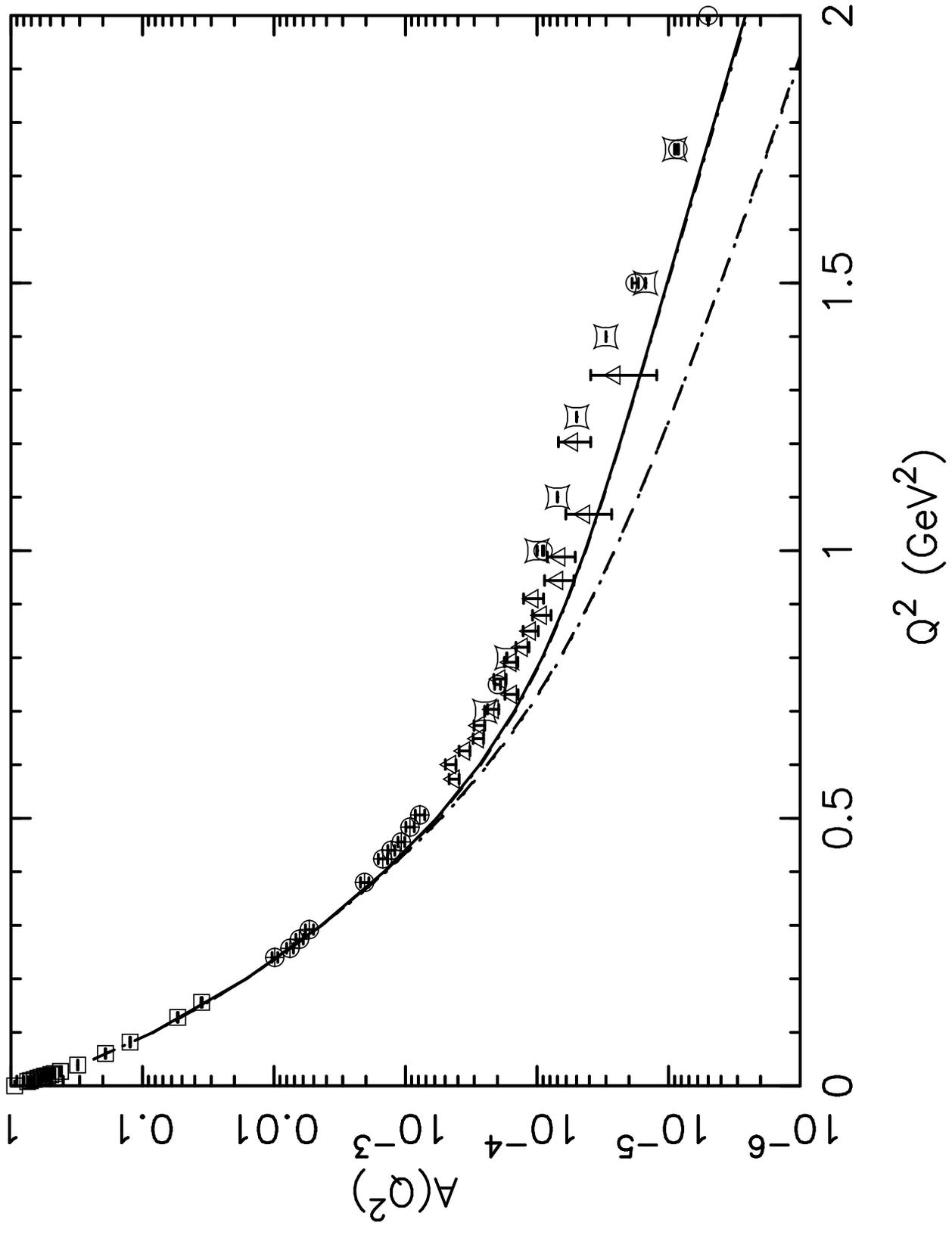
\begin{figure}
\caption{$A(Q^2)$ for the Argonne potential with G-K (solid) and MMD (dashed) form factors, and
for the Reid potential with G-K (dash-dot) and MMD (dotted) form factors.  The data come from
Refs. [41] (squares), [42] (crossed circles), [43] (triangles), [44] (open circles) and [56] (curved squares).}
\label{paapic2plot}
\end{figure}

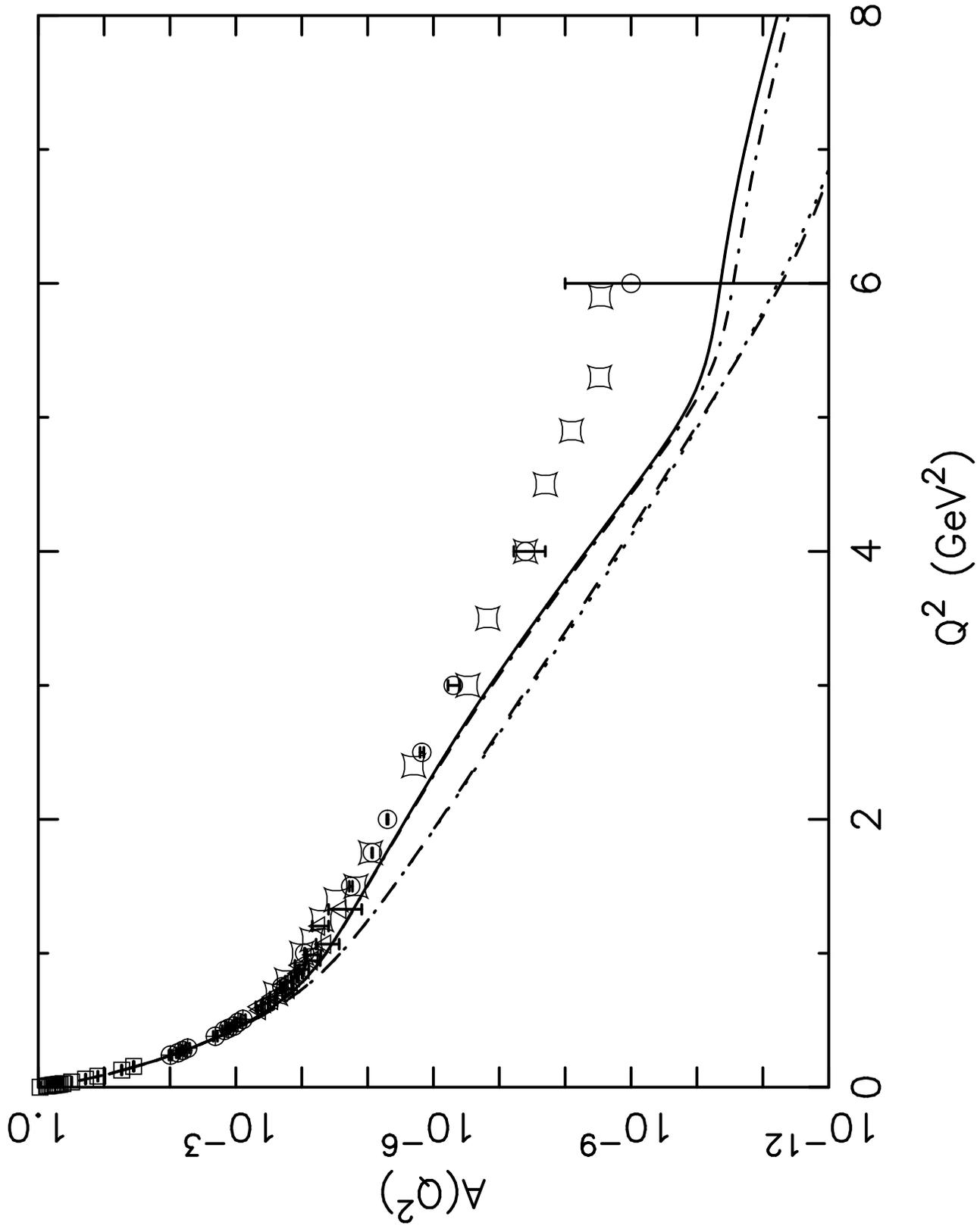
\begin{figure}
\caption{$A(Q^2)$ for the Argonne potential with G-K (solid) and MMD (dashed) form factors, and
for the Reid potential with G-K (dash-dot) and MMD (dotted) form factors.  The legend for the data
is the same as in Figure 7.}
\label{paapic8plot}
\end{figure}

Figure \ref{paapic2plot}
%apic2plot.ps
displays the results for $A(Q^2)$ up to 2 ${\rm GeV}^2$ for both potentials and both
form factor parameterizations; Figure \ref{paapic8plot}
%apic8plot.ps
extends the calculations to 8 ${\rm GeV}^2$.  The data come from Refs.\ \cite{simon,galst,elias,arn75}.
Differences among the various calculations begin to appear at intermediate momentum transfers.  For
$A(Q^2)$, with $Q^2$ between 0.5 and 3 GeV${}^2$, the PFSA combined with the Gari-Kr\"umpelmann form
factors fit the data fairly closely, while the Mergell-Meissner-Drechsel form factors produce results
that fall short by an order of magnitude.  This pattern occurs in other impulse and spectator calculations as
well.  In the front-form calculations of Chung {\it et al}.\  \cite{cckp}, the fit for
an earlier G-K parameterization (using the Argonne $v_{14}$ potential) is even better, while the
H\"ohler (on which the newer MMD form factors were based) calculations again fall an order of magnitude
short.  In Lev {\it et al}.\  \cite{lev} the two curves are closer, though the G-K remains higher
and fits the data out to 2.0 GeV${}^2$.  In the nonrelativistic calculations of Carlson and
Schiavilla \cite{carls}, which cover the range 0--2.4 GeV${}^2$ and use only the H\"ohler form
factors, impulse approximations using various potentials (including the Argonne $v_{18}$) all fall
nearly an order of magnitude short in the intermediate range.  In the work of Van Orden {\it et al}.\ 
\cite{wally}, which contains an impulse approximation that fits most of the data for all three form factors quite
closely, the variation from the $A(Q^2)$ data starts at 2.0 GeV${}^2$ and is an order of magnitude
short at high (8 GeV${}^2$) momentum transfers.

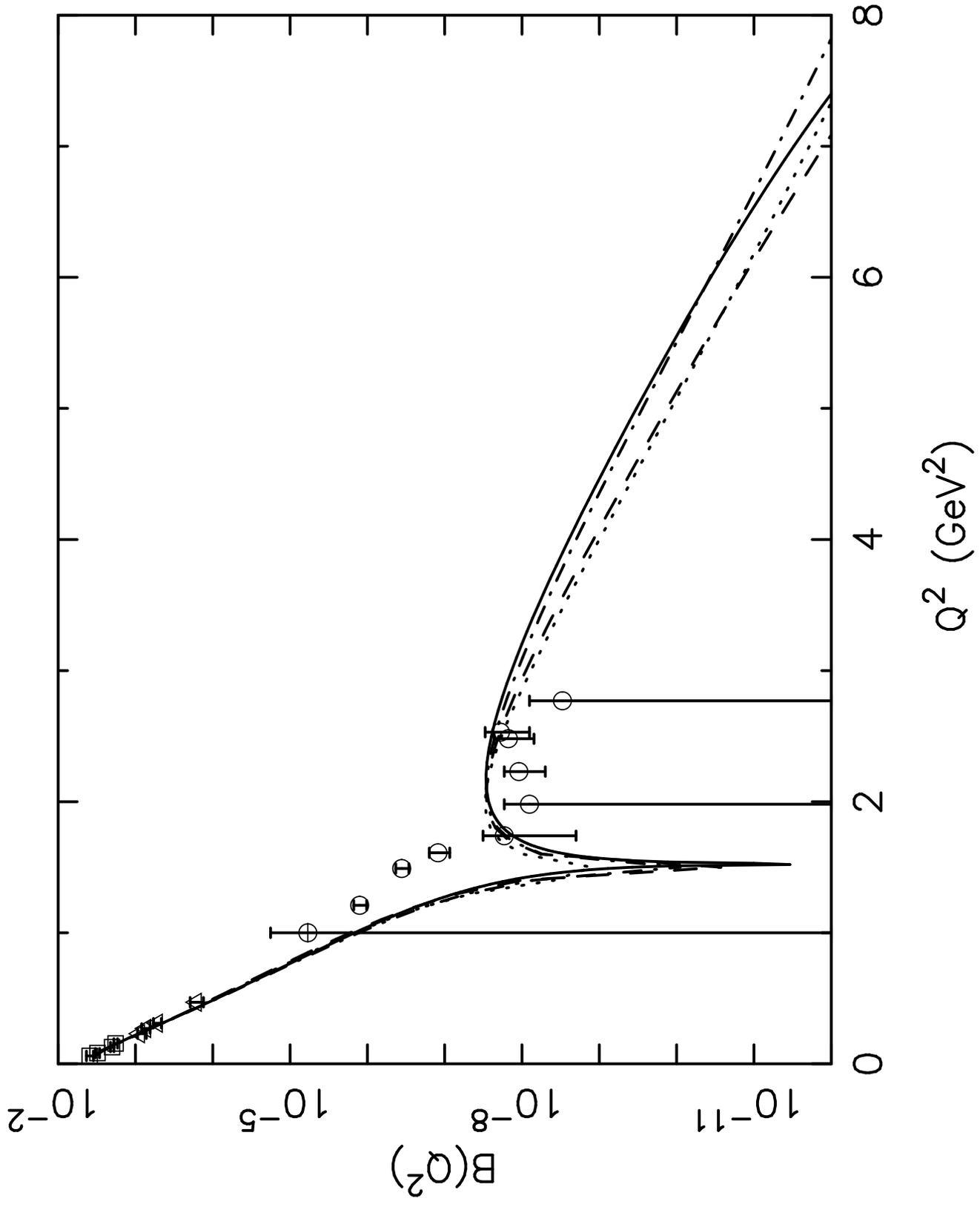
\begin{figure}
\caption{$B(Q^2)$ for the Argonne potential with G-K (solid) and MMD (dashed) form factors, and
for the Reid potential with G-K (dash-dot) and MMD (dotted) form factors.  The data come from Refs.
[41] (squares), [45] (triangles), [46] (crossed circle), and [47] (open circles).}
\label{pabpic8plot}
\end{figure}

The elastic observable $B(Q^2)$, related directly to the magnetic form factor $G_M$, is displayed
in Figure \ref{pabpic8plot}
%bpic8plot.ps
up to 8 ${\rm GeV}^2$ using both potentials and both sets of parameterizations.
The data come from Refs.\ \cite{simon,bucha,marti,arn87}.
The PFSA calculation of $B(Q^2)$ in the intermediate region 0.5--3 GeV${}^2$, a region which contains
all presently available data, fits that data poorly, though the differences between the two form factor
parameterizations are less marked.  Both parameterizations exhibit a zero at 1.5$\pm$0.1 GeV${}^2$,
causing a wide discrepancy with experiment; data suggest a zero nearer 1.9 GeV${}^2$.  Again Chung {\it et al}.\ 
and Van Orden {\it et al}.\  fit the $B(Q^2)$ data quite well.  In contrast, Chung {\it et al}., using the Paris and the Bonn wavefunctions
instead of the Argonne $v_{14}$, produce results quite similar to the point form's.  Carlson and Schiavilla
obtain a zero at 2.2 GeV${}^2$ only with the Nijmegen potential; their other potentials
reproduce the zero at 1.6 GeV${}^2$.  Lev {\it et al}.\  produce zeroes between 1--2 GeV${}^2$ in the 0--4 GeV${}^2$
region using various potentials and parameterizations.  The position of the zero in all these forms seems to be
the most salient feature of calculations of $B(Q^2)$, affecting as it does the deviation from data in the
1--2 GeV${}^2$ range.

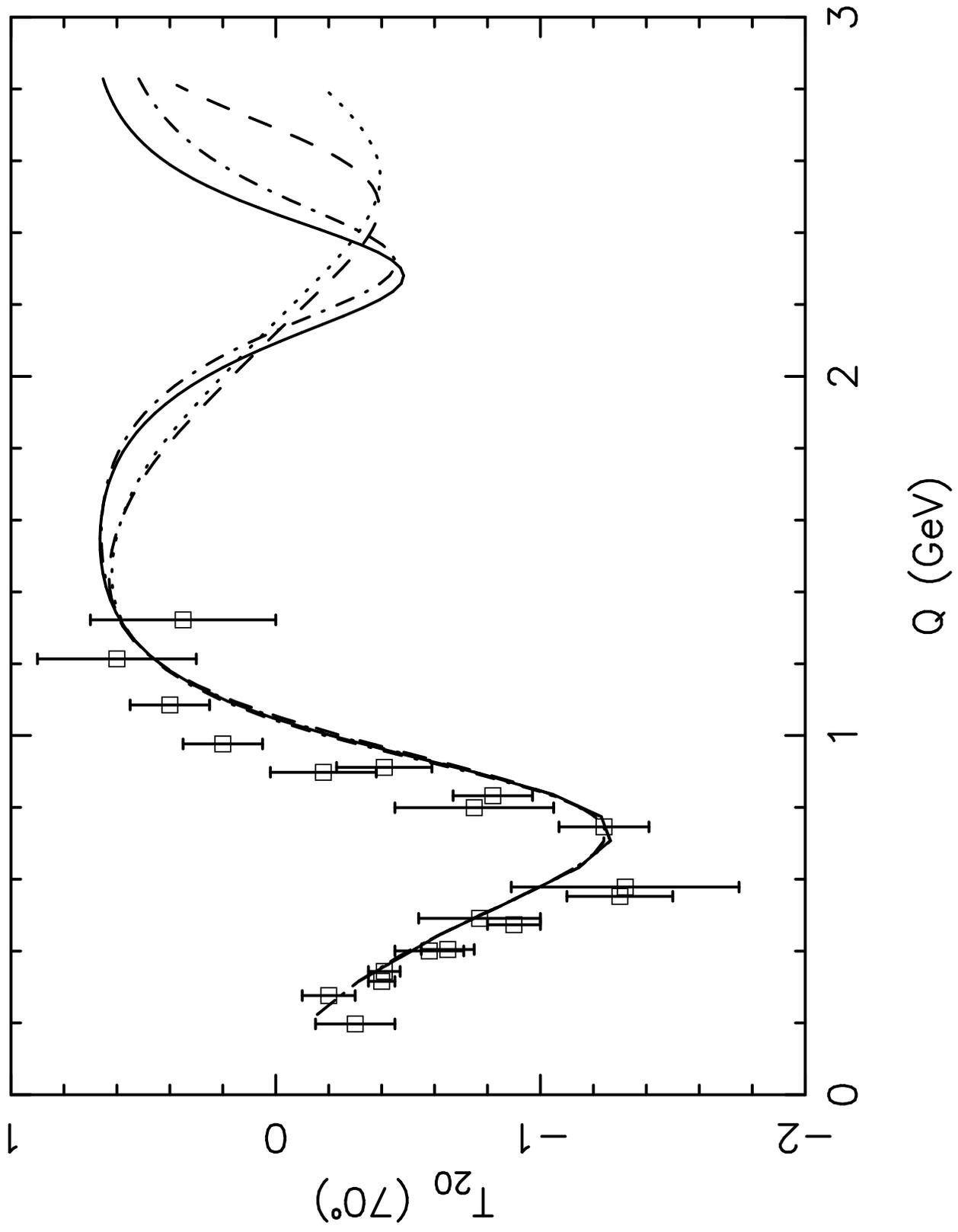
\begin{figure}
\caption{$T_{20}(Q^2)$ for the Argonne potential with G-K (solid) and MMD (dashed) form factors, and
for the Reid potential with G-K (dash-dot) and MMD (dotted) form factors.  The data are
compiled from Refs. [48-55].}
\label{pat20pic8plot}
\end{figure}

The tensor polarization, $T_{20}(Q^2)$, is displayed in Figure \ref{pat20pic8plot}, with data from Refs.\ 
\cite{schul,gilma,the,garco,ferro,dmitr,bouwh,abbot}.
%t20pic8plot.ps
As with $B(Q^2)$, measurements of $T_{20}(Q^2)$ have only investigated the low (0--0.5 GeV) and intermediate
(0.5--1.5 GeV) ranges of momentum transfer.  In both ranges, the MMD and G-K parameterizations produce
identical results in the PFSA.  In the intermediate range, the results fall slightly below the data.  The impulse approximations of Carlson and Schiavilla
as well as Lev {\it et al}.\  do this too, while Van Orden {\it et al}.\  and Chung {\it et al}.\  produce curves
that fit modern data quite closely.

Finally, it is instructive to compare the results obtained in point-form dynamics to the results
of nonrelativistic impulse calculations to get some idea of the nature of the relativistic effects; and
to compare the results in the point form to the same relativistic calculations done assuming
$|(p'_1-p_1)^2|=Q^2$ (that is, pulling the nucleon form factors outside of the integral)
to examine how the point-form momentum transfer affects the results.

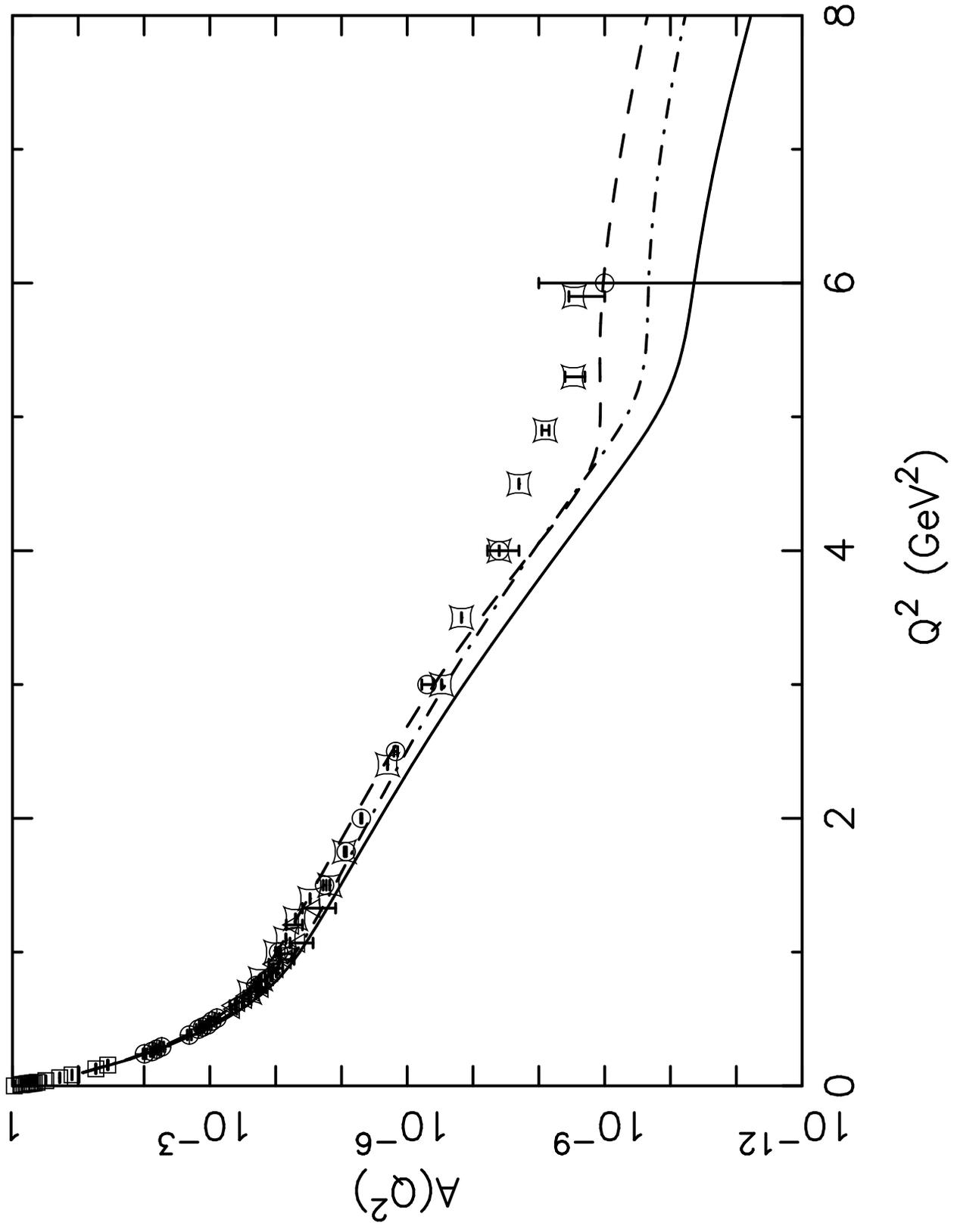
\begin{figure}
\caption{Point-form (solid), nonrelativistic (dashed), and constant-$Q^2$ (dash-dot)
results for $A(Q^2)$ using the Argonne potential and the G-K form factors.}
\label{panonapic8plot}
\end{figure}

The results (using the Argonne $v_{18}$ potential and the Gari-Kr\"umpelmann form factors)
for $A(Q^2)$ are displayed in Figure \ref{panonapic8plot}.
%nonapic8plot.ps
For low momentum transfers, all three agree.  In the range of 1.0--5.0 GeV${}^2$, the nonrelativistic
and PFSA calculations decrease similarly, the nonrelativistic curve consistently higher.  The constant-$Q^2$
calculation in this region gradually rises from the PFSA to the nonrelativistic curve.  Above 5.0 GeV${}^2$,
all three curves systematically decrease, nonrelativistic above constant-$Q^2$ above PFSA.
While a relativistic treatment is needed as a matter of principle at high momentum transfers,
it is clear that in these calculations the effects of combining the PFSA with point-form
quantum models has a tendency to reduce the structure function $A(Q^2)$ at high $Q^2$.
One clear cause of this is that
\begin{equation}
|(p'_i-p_i)^2|>Q^2;
\label{pfbigq2}
\end{equation}
the magnitude of the point-form momentum transfer is greater than the magnitude of the nonrelativistic
momentum transfer.  Because the form factors depend on the magnitude of the momentum transfer, they
therefore drop off more quickly in the point form.  This reduces the point-form results
in comparison to the constant-$Q^2$ calculations, as Figure \ref{panonapic8plot} shows:
as the momentum transfer increases, the two calculations diverge further from each other.

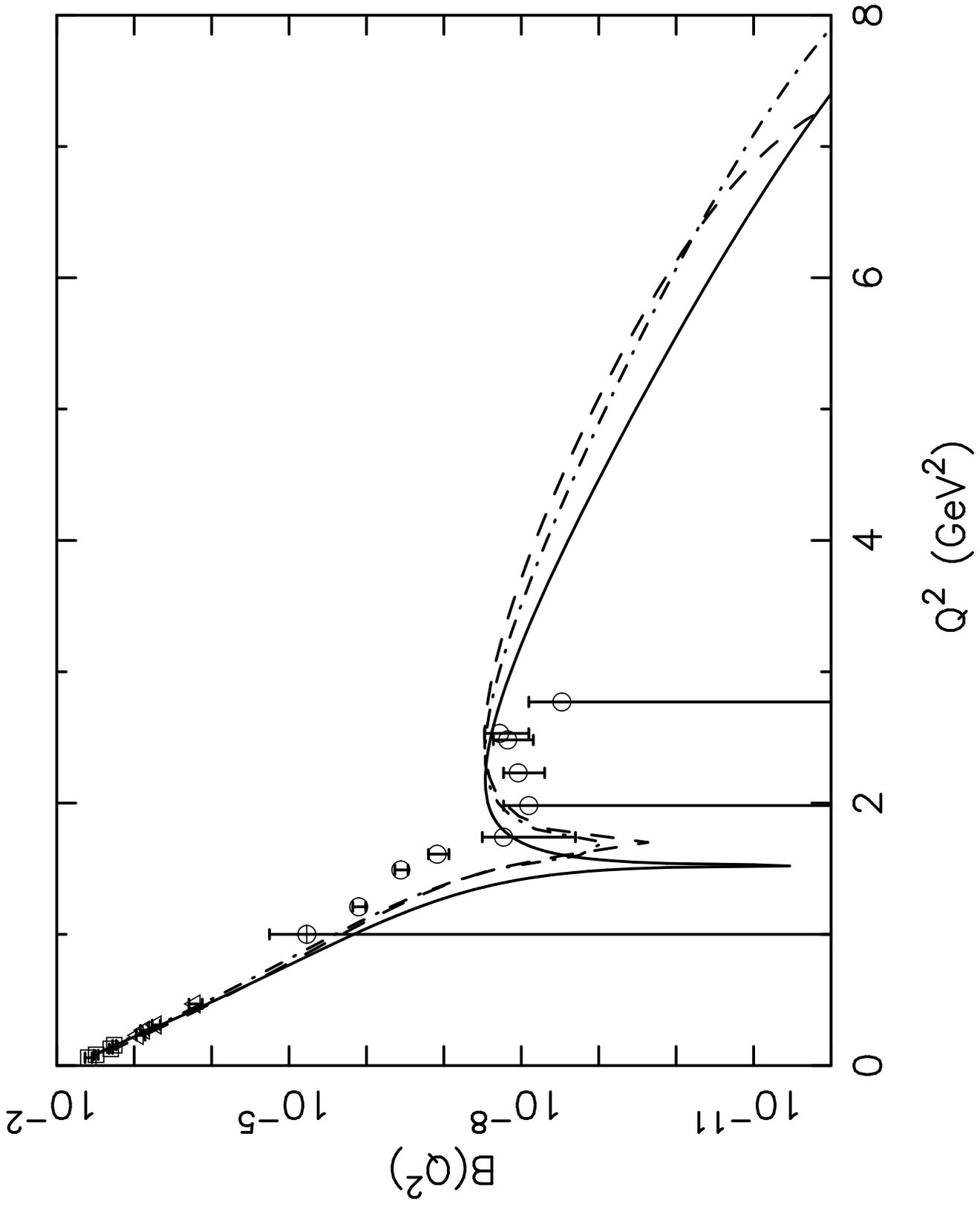
\begin{figure}
\caption{Point-form (solid), nonrelativistic (dashed), and constant-$Q^2$ (dash-dot)
results for $B(Q^2)$ using the Argonne potential and the G-K form factors.}
\label{panonbpic8plot}
\end{figure}

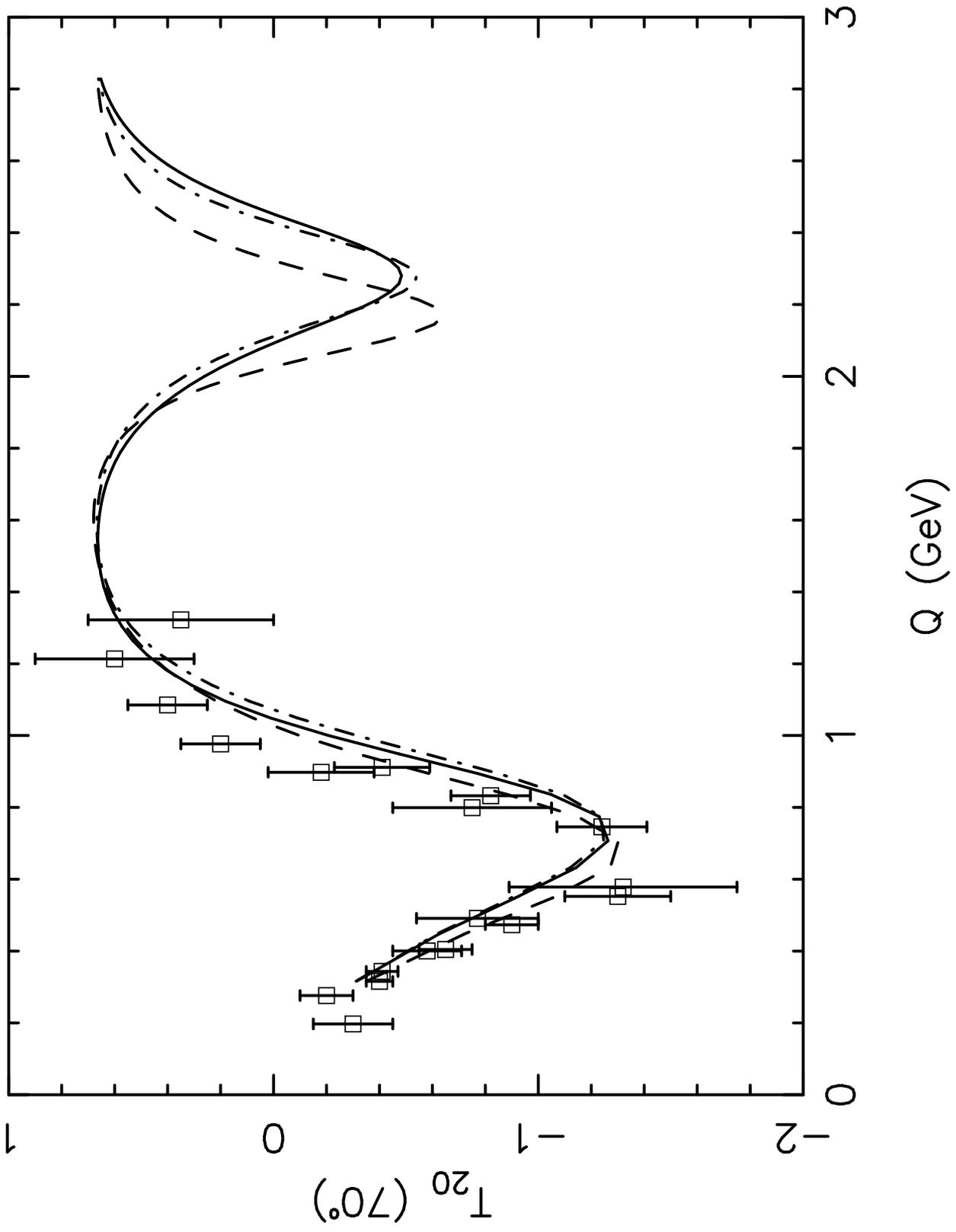
\begin{figure}
\caption{Point-form (solid), nonrelativistic (dashed), and constant-$Q^2$ (dash-dot)
results for $T_{20}(Q^2)$ using the Argonne potential and the G-K form factors.}
\label{panont20pic8plot}
\end{figure}

The point-form and nonrelativistic results for $B(Q^2)$ and $T_{20}(Q^2)$ do not exhibit as dramatic
differences as they did for $A(Q^2)$.  Again in the graphs of $B(Q^2)$ and $T_{20}(Q^2)$,
Figures \ref{panonbpic8plot} and \ref{panont20pic8plot},
%nonbpic8plot.ps
one sees the similar but increasingly divergent results obtained from the constant-$Q^2$ and the
PFSA methods.
%nont20pic8plot.ps

\section{Conclusion}

This work has used the point form of relativistic dynamics to calculate elastic deuteron
form factors.  The point form stands somewhat between the covariant approaches and direct-interaction
approaches mentioned in the introduction in that it is on the one hand manifestly covariant
(because the Lorentz generators are kinematic) and it is the mass that is ``off-shell''
(rather than the energy as is the case with the instant form.)  On the other hand the point form
is one of the forms of dynamics listed by Dirac, in which all of the interactions are in the
four-momentum generators.  Moreover there is a natural way in which one-body currents can be
introduced in the point form (called the point-form spectator approximation) that satisfies the
correct Poincar\'e and charge conservation properties.

We have shown that the PFSA produces results consistent with other impulse and spectator approximations.
Within the range Schiavilla and Riska examined, for example, their impulse approximation and the PFSA
(using G-K form factors) predict nearly identical results for $A(Q^2)$ and $T_{20}(Q^2)$; and though the zero they predict for $B(Q^2)$
falls near 2.0 GeV${}^2$ rather than the PFSA's 1.6, the fall-off from the data begins near
0.5 GeV${}^2$ in both.  The calculations of Kobushkin and Syamtomov \cite{kobus} (before using their
approach of reduced transition amplitudes) and the PFSA (G-K) results for $A(Q^2)$ and $B(Q^2)$ nearly
duplicate each other, as do the results for $T_{20}(Q^2)$ except in the high-momentum range, where no data
are currently available.  And although the calculations of Chung {\it et al}.\  with earlier G-K form factors
and the Argonne $v_{14}$ potential fall quite close to the data for all three observables, their calculations using
H\"ohler \cite{hoehl} form factors and potentials show the same salient points:  the deviation from data beginning in $A(Q^2)$ near
1 GeV${}^2$ and in $B(Q^2)$ near 0.5 GeV${}^2$; the location of the first zero in $B(Q^2)$ between 1.5--2.0 GeV${}^2$;
and the first minimum in $T_{20}(Q^2)$ around 0.8 GeV${}^2$.  The work of Van Orden {\it et al}.\  gives
results similar to Chung {\it et al}., except that the high-$Q^2$ behavior of $A(Q^2)$ and $B(Q^2)$ is nearly level at $10^{-8}$,
while Chung {\it et al}.\  and the PFSA show gradual decreases at $10^{-11}$ for $A(Q^2)$ and at $10^{-8}$ for $B(Q^2)$.
In contrast, Lev {\it et al}.\  calculate
that the H\"ohler form factors produce results for $A(Q^2)$ that lie close to the data, while the G-K
form factors fit $A(Q^2)$ up to 2 GeV${}^2$ but produce results increasingly higher than the data thereafter.
Their results for $B(Q^2)$ and $T_{20}(Q^2)$ are similar to those of Van Orden {\it et al}.\  and Chung {\it et al}. 

This work also addressed the sensitivity of PFSA results to different nucleon-nucleon interactions
and different parameterizations of the nucleon form factors.  In almost every instance it was found
that the two nucleon-nucleon potentials produced only slight, if any, differences in the form factors
and elastic observables.  This may not be surprising considering that the Argonne $v_{18}$ and Reid '93
nucleon-nucleon interactions produce nearly identical momentum-space wavefunctions on the momentum scale of interest.

Much more pronounced were the differences between the Gari-Kr\"umpelmann and the Mergell-Meissner-Drechsel
parameterizations of the nucleon form factors.  As the momentum transfers become higher, the two often
predict significantly different results.  These are most notable in the deuteron form factors
$G_E$ and $G_Q$, which are sensitive to the neutron electric form factor; the G-K parameterization,
whose neutron electric form factor falls off markedly more rapidly than the MMD, produces deuteron form
factor zeroes in the intermediate range that occur at higher momentum transfers in the MMD results.
That this phenomenon is due to the neutron electric form factor is supported by the similarity of results
for $G_M$, where the nucleon magnetic form factors dominate the calculations.

Aside from differences due to varying potentials and nucleon form factors, the consequences of the
point form's nontrivial momentum transfer have also been examined.  We have shown that the point-form
momentum transferred to a nucleon is greater than the $Q^2$ transferred to the deuteron, and that
its deviation increases with increasing $Q^2$.  This results in a lowering of the deuteron form
factors and elastic scattering observables compared to nonrelativistic calculations.  The greater
magnitude of the point-form momentum transfer causes a quicker fall-off of the nucleon form factors,
and some deviation from nonrelativistic calculations was attributed to this.

Additional dynamically consistent two-body currents must be added to the PFSA in order to bring
the calculations into agreement with data at intermediate to high momentum transfers.  Such currents
have not been considered in this preliminary work; but other calculations that include dynamical
two-body currents (see references \cite{humme,lev,cckp,schia}) suggest that the addition of dynamical
currents is capable of reconciling the differences between various impulse or spectator approximations
with data.

In the point form it is quite easy to interpret Feynman diagrams for nucleon-nucleon
scattering with the production of a photon as a current matrix element satisfying the requirements
given in Section III.
However, while the addition of such current matrix elements may produce better agreement with data,
it does not provide a systematic procedure for constructing two-body currents.  What is needed is a
procedure for constructing conserved currents from one-body currents and the dynamical mass operator.
Models based on such a procedure are being developed.

\appendix
\section{Point-form Momentum Transfer}

The momentum transfer $(p'_i-p_i)$ seen by nucleon $i$ can be computed following Ref.\ \cite{klink}.
Suppose that the momentum transfer is along the z-axis.
\begin{eqnarray}
B(v_{in}) &=& \left( \begin{array}{cccc}
  \cosh\Delta/2 & 0 & 0 & \sinh\Delta/2 \\
  0 & 1 & 0 & 0 \\
  0 & 0 & 1 & 0 \\
  \sinh\Delta/2 & 0 & 0 & \cosh\Delta/2 \end{array} \right), \\
B(v_{out})&=&\left( \begin{array}{cccc}
  \cosh\Delta/2 & 0 & 0 & -\sinh\Delta/2 \\
  0 & 1 & 0 & 0 \\ 0 & 0 & 1 & 0 \\
  -\sinh\Delta/2 & 0 & 0 & \cosh\Delta/2 \end{array} \right),
\end{eqnarray}
are the boosts that take the deuteron from the center of momentum frame to the Breit frame
(where ${\bf P}'_{\rm tot}=-{\bf P}_{\rm tot}$.)  For elastic scattering,
\begin{equation}
\sinh\Delta/2=\sqrt{Q^2\over 4m^2_D}.
\end{equation}
The initial energies and momenta are then:
\begin{eqnarray}
E_1 &=& \omega\cosh\Delta/2+k_z\sinh\Delta/2; \nonumber \\
p_{1z} &=& k_z\cosh\Delta/2+\omega\sinh\Delta/2; \nonumber \\
E_2&=&\omega\cosh\Delta/2-k_z\sinh\Delta/2; \nonumber \\
p_{2z}&=&-k_z\cosh\Delta/2+\omega\sinh\Delta/2;
\end{eqnarray}
where $\omega$ and ${\bf k}$ are center of momentum variables.

In this notation, $k_z$ refers to the relative z-axis momentum of particle one.  This
convention gives rise to the following relations:
\begin{eqnarray}
\omega'&=&\omega\cosh\Delta\mp k_z\sinh\Delta; \\
k'_z &=&k_z\cosh\Delta\mp\omega\sinh\Delta;
\end{eqnarray}
where the minus signs are used when particle one is struck, the plus signs when particle two is struck.
Suppose for illustration that particle one is struck.  The final energies and momenta will then be
\begin{eqnarray}
E'_1&=&\omega\cosh 3\Delta/2 - k_z\sinh 3\Delta/2; \nonumber \\
p'_{1z}&=&k_z\cosh 3\Delta/2-\omega\sinh 3\Delta/2; \nonumber \\
&& E'_2=E_2; \qquad p'_{2z}=p_{2z}.
\end{eqnarray}
Now some hyperbolic trigonometry reveals that
\begin{equation}
(p'_1-p_1)^2=4(k^2_z-\omega^2)\sinh^2\Delta.
\end{equation}
Since
\begin{equation}
\sinh\Delta = 2\sqrt{Q^2\over 4m^2_D}\sqrt{1+{Q^2\over 4m^2_D}},
\end{equation}
and
\begin{equation}
k^2_z-\omega^2=k^2_z-m^2_n-{\bf k}^2=-(m^2_N+{\bf k}^2_{\bot}),
\end{equation}
the resulting Equation \ref{biggerq2} is established.
\pagebreak

\begin{table}
\begin{center}
\begin{tabular}{|c|c||c|c|c|c|c|}
Moment & Units & Experimental & Argonne $v_{18}$ & Argonne $v_{18}$ & Reid '93 & Reid '93 \\ \hline
 & & & G-K & MMD & G-K & MMD \\ \hline
$\mu_D$ & $e\mu_N$ & 0.85741 & 0.8613 & 0.8623 & 0.8615 & 0.8625 \\ \hline
$Q_D$ & $e/{\rm GeV}^2$ & 7.3422 & 6.6 & 6.6 & 6.6 & 6.6 \\ \hline
\end{tabular}
\end{center}
\caption{The magnetic dipole and electric quadrupole moments, computed as $Q^2\rightarrow 0$.}
\label{static}
\end{table}

\pagebreak
\mediumtext

\end{document}